\documentclass[10pt]{iopart}
\usepackage{iopams}
\usepackage{cite}
\usepackage{multirow}
\usepackage[section]{placeins}
\usepackage{amssymb}
\usepackage{stfloats}
\usepackage{textcomp}
\usepackage{gensymb}
\usepackage{xpatch}
\usepackage{tabularx}

\usepackage[pdftex]{graphicx}
    \DeclareGraphicsExtensions{.pdf,.png}
\usepackage{dcolumn}
\usepackage{bm}
\usepackage{float}
\usepackage{color}

\begin{document} 
\title[Stability/accuracy conditions of PIC/MCC]{{\color{black} Revisiting  the  numerical  stability/accuracy  conditions  of  explicit  PIC/MCC simulations of low-temperature gas discharges}}

\author{M.~Vass$^{1,2}$, P.~Palla$^{3,4}$, P. Hartmann$^2$}

\address{$^1$ Department of Electrical Engineering and Information Science, Ruhr-University Bochum, 44780 Bochum, Germany}
\address{$^2$ Institute for Solid State Physics and Optics, Wigner Research Centre for Physics, 1121 Budapest, Konkoly-Thege Mikl\'os str. 29-33, Hungary}
\address{$^3$ Faculty of Science, E\"otv\"os Lor\'and University, 1117 Budapest, P\'azm\'any P\'eter s\'et\'any 1/A, Hungary}
\address{$^4$ Robert Bosch Kft., 1103 Budapest, Gy\"omr\H{o}i \'ut 104., Hungary}

\ead{vass@aept.ruhr-uni-bochum.de}

\begin{abstract}
Particle-in-cell with Monte Carlo collisions (PIC/MCC) is a fully kinetic, particle based numerical simulation method with increasing popularity in the field of low temperature gas discharge physics. Already in its simplest form (electrostatic, one-dimensional geometry, and explicit time integration), it can properly describe a wide variety of complex, non-local, non-linear phenomena in electrical gas discharges at the microscopic level with high accuracy. However, being a numerical model working with discretized temporal and (partially) spatial coordinates, its stable and accurate operation largely depends on the choice of several model parameters. Starting from four selected base cases of capacitively coupled, radio frequency driven argon discharges, representing low and intermediate pressure and voltage situations, we discuss the effect of the variation of a set of simulation parameters on the plasma density distribution and the electron energy probability function. The simulation parameters include the temporal and spatial resolution, the PIC superparticle weight factor, as well as the electron reflection and the ion-induced electron emission coefficients, characterizing plasma-surface interactions. 
\end{abstract}

\maketitle
\ioptwocol

\section{Introduction}\label{sec:intro}

Low-pressure gas discharges, besides being fundamentally interesting nonlinear complex physical systems, have become irreplaceable tools for numerous modern technologies. The fabrication of microelectronics, thin films, photovoltaic panels, as well as the operation of gas lasers, the plasma treatment of special-purpose surfaces, and the conversion of harmful gases into neutral compounds are just a few examples to mention \cite{Fridman,Lieberman,Adamovich2017,decadal2021}. In many cases, low gas pressure is desired to enable the acceleration of ions to high energies due to the limited number of collisions with the neutral background gas. In such cases, only models that include non-local particle transport are assumed to provide reliable descriptions due to the long mean free path of the electrons and ions in the plasma. Particle-based methods bear the advantage of relying mostly on first principles, adopting fewer approximations (compared to continuum methods) at the cost of requiring detailed microscopic information on the elementary collision processes between the constituent particles, as well as high computational demand. In the field of radio-frequency (RF) driven technological plasmas the particle-in-cell method, complemented with the Monte Carlo type of binary collision model (PIC/MCC) is the most popular particle based approach to date \cite{Tskhakaya2007,Alves18, Brieda}.

In principle, particle-based simulations should trace the motion of all microscopic constituents of the systems together with all (short-, and long-range) interactions among the particles, the external fields, and the plasma-wall interactions. This is, unfortunately, technically not possible due to the vast number of particles and interactions. The way this problem is approached in PIC/MCC simulations is that (i) an ensemble of real microscopic particles that share a small volume in the phase-space are grouped and replaced by a single superparticle, and (ii) the pairwise interaction between the charged particles is simplified to a mean-field approach where the charge-charge interaction is mediated by the self-consistent electromagnetic fields combining the contributions of both external and internal charges and fields \cite{Birdsall_2004}. In the case of the electrostatic PIC/MCC model, the subject of the present study, the Poisson equation is solved for the time-dependent electric field on a numerical grid using the finite difference scheme. In practice, the reduction of the particle count is realized with the introduction of an arbitrary weight factor defining the number of elementary particles represented by each superparticle. The weight factor can be chosen to be constant and equal for all species, but more sophisticated concepts are possible as well \cite{Welch07,Sun16}. 

The superparticles are traced in continuous phase space but using discretized time, advancing step by step. The solution of the equation of motion can be performed using explicit or implicit formulations and applying iterative methods of any order. The most straightforward, and thus most common, approach is based on the explicit leapfrog integration, which belongs to the family of second-order symplectic methods \cite{Birdsall_2004}.

Every implementation of the PIC/MCC simulation, similarly to any other numerical solution method, can provide solutions that approximate the exact solution of the target equation by some finite level of accuracy, depending on the choices of several numerical parameters, like the size of the timestep or the spatial resolution of the numerical grid. Ideally, one should be able to estimate these numerical parameters once an expected level of accuracy is defined. Unfortunately, this is only possible in some very limited numbers of simple equations. The choice of the numerical parameters does, however, not only influence the accuracy of the computations but can lead to instabilities of the applied method, resulting in non-physical oscillations or divergence of the computation \cite{Birdsall_2004}. To guarantee the proper operation of numerical solution methods a set of {\it stability criteria} has to be fulfilled and furthermore, usually an even stricter set of {\it accuracy criteria} should be taken into account to provide stable and accurate solutions.

Of course, even if the numerical scheme is capable of solving the physical model and properly chosen numerical parameters ensure stable and accurate operation of the simulation, this still does not guarantee that the model itself accurately approximates the real physical system. Ideally, all simulation models and codes should be validated against experimental data, which, however, is not possible in many cases. As an alternative to experimental verification, the cross-benchmarking of simulations developed independently by different groups should be performed. The issues of code verification and validation have been addressed in previous studies~\cite{turner2013simulation,Turner2016,turner2017computer,Riva17,kilian2017,Carlsson2016,OConnor2021,Fierro21}.

In the case of electrostatic, explicit PIC/MCC simulations the traditional view, based on recent publications, is that the following 5 stability criteria have to be checked before accepting the results of the computations (see e.g. \cite{kim2005particle,lymberopoulos1995two}).

\begin{enumerate}
    \item {\it The spatial grid has to resolve the electron Debye-length.} $\lambda_{\rm D}$ is the natural length scale of the plasma. Important physical phenomena, like the screening of excess charges, take place on this scale. The lack of spatial resolution will lead to an improper description of the underlying physics.
    \item {\it The integration timestep of the equations of motion has to be small enough to ensure that trajectories are computed accurately.} The electron plasma frequency, $\omega_{\rm p,e}$, defines a characteristic time-scale at which the lightest and fastest plasma constituents (electrons) can respond to variations of the force-field. Missing to resolve particle motion on this time scale can easily result in nonphysical trajectories and thus, inaccurate results.
    \item {\it The collision probabilities in each timestep should be sufficiently small.} During the Monte Carlo type of collision management, each particle has a finite chance to participate in collisions in each timestep, and random numbers are generated to decide about individual events. In case of an improper choice of the timestep, a chance of multiple collisions happening in one timestep can be large and events can be missed during the simulations.
    \item {\it Particles should not fly a longer distance during a timestep than the grid division, as expressed by the Courant–Friedrichs–Lewy (CFL) condition} \cite{karimabadi2005new}. This condition was originally derived for the numerical solution of partial differential equations \cite{Courant67} with applications in computational fluid dynamics \cite{Laney98}, but seems relevant to PIC/MCC simulations as the force-field values are computed on the numerical grid. For the particles to respond accurately to the fields it is important to properly sample the local force values along the trajectories. 
    \item {\it Accurate results require a high number of particles per grid cell.} Artificial numerical heating \cite{turner2006kinetic} appears in the system due to statistical fluctuations, which are stronger when a smaller number of superparticles is used. Ideally, the results should not depend on the particle number (or on the weight factor), however, this is not exactly the case, see e.g. \cite{Erden,Sun16}.  
\end{enumerate}

The last condition in the list represents a real bottleneck in the PIC/MCC approach, as the choice of the number of particles represents a compromise between accuracy and the execution time of the computer program.

It needs to be mentioned that, in the case of low-pressure electrical gas discharges, none of the above criteria can be formulated rigorously providing exact threshold values, and serve mostly as general rules that are intended to help the researcher make decisions. For example, the first two criteria rely on the Debye-length and the electron plasma frequency. Strictly, both quantities are derived in the frame of first order (linear) approximation for plasmas in thermal equilibrium \cite{Lieberman}, which is certainly not valid here. Similarly, the original derivation for the CFL-condition assumed a PDE with spatial as well as temporal derivatives \cite{Courant67}. As in the electrostatic approximation only Poisson's equation is solved on the numerical grid, it is not immediately obvious whether the same criterion applies here.

In this work, using our 1d3v (one dimensional in real space and three dimensional in velocity space) electrostatic PIC/MCC simulations, we present a systematic study on the effect of a set of model parameters on the principal discharge characteristics of capacitively coupled RF argon discharges driven at 13.56 MHz in different operating regimes. The numerical parameters of interest are: (i) the time-resolution, (ii) the spatial resolution, (iii) the superparticle weight factor $W$, as well as parameters characterizing surface processes: (iv) the ion induced electron emission yield $\gamma$, and (v) the elastic electron reflection coefficient $R$. To illustrate the parameter dependence, the electron density profiles, and bulk electron energy probability function will be presented, being the two most sensitive and most relevant quantities for many applications.

\section{Computational method}

Our simulation code is described in great detail in \cite{eduPIC}, therefore, here we just briefly list the main features to provide an overview. 

Each period of the driving voltage waveform $V(t) = \phi_0 \cos(2\pi f t)$, with a constant frequency of $f=13.56$~MHz, quantified by the voltage amplitude $\phi_0$, is divided into $N_t$ individual timesteps with duration $\Delta t = 1/(f N_t)$. The traditional six basic steps of the iterative PIC/MCC cycle, executed in order in every timestep, are implemented as:
\begin{enumerate}
    \item {\it Computing the charged particle densities:} A linear particle shape function with a total width of $2\Delta x$ is used to assign the contribution of each particle (with coordinates $x_{\rm i}$ for the $i$-th particle) to the numerical grid. The grid spacing is $\Delta x = L/(N_x-1)$, where $L$ is the length of the discharge gap and $N_x$ is the number of grid points.
    \item {\it Computing the potential and electric field:} With the charge density distribution, $\rho(x,t)$ and the Dirichlet boundary conditions ($\phi_{(x=0)}=V(t)$ and $\phi_{(x=L)}=0$) at hand the electrostatic potential distribution, $\phi(x,t)$ is obtained by solving Poisson's equation using the tridiagonal Thomas algorithm \cite{Thomas1949}. The electric field distribution, $E(x,t)$ is obtained by numerical differentiation of the potential.
    \item {\it Computing the forces acting on the particles:} From the electric field distribution, available on the numerical grid, the field values at the position of the particles are obtained by linear interpolation.
    \item {\it Moving the particles:} The explicit leap-frog time-integrator is used to advance the particle position and to update the particle velocity vectors.
    \item {\it Adding/removing particles at the boundaries:} Particles with their newly updated position being outside the simulation region are either removed from the simulation or, depending on the surface parameters defined in the simulation, can participate in surface processes. In the case of electrons, elastic reflection is implemented with a probability defined by the reflection coefficient, $R$. In the case of Ar$^+$ ions, the electron emission yield, $\gamma$ defines the probability of a low energy electron being emitted from the surface as a consequence of each individual ion impact. The decision for each individual event is made using uniformly distributed pseudo random numbers obtained with the Mersenne Twister 19937 generator algorithm \cite{Mersenne}.
    \item {\it Checking and executing collisions:} The probability of collisions is evaluated for every particle in every timestep based on the set of energy dependent cross section data. For electron - Ar collisions, elastic scattering, excitation, and ionization processes are implemented using data from \cite{phelps1999cold}. In the case of Ar$^+$ - Ar collision, isotropic elastic and symmetric charge transfer processes are taken into account as suggested in \cite{phelps1994application}. During this step, pseudo random numbers are used to decide about the occurrence of collisions, the selection of the process type, and for the assignment of post-collision particle velocities, all based on physically relevant statistical distributions \cite{eduPIC}.
\end{enumerate}

This PIC/MCC cycle is repeated until convergence is reached, typically lasting a few thousand RF periods. Measurements are started after this initial simulation phase. During the measurements, only statistical fluctuations and no systematic drift was observed in every quantity.

\section{Results}

In this section, we present a systematic study of the parameter space of the numerical input parameters of 1d3v PIC/MCC simulations. To investigate physically different conditions, we have chosen four ``base'' cases. These four cases correspond to four ``physical limits'': low/high pressure with low/high voltage amplitude, respectively, excited by a single frequency waveform of a given frequency. The effect of a given numerical parameter will be investigated for all ``base'' cases simultaneously.

\begin{table*}[ht]
    \caption{Simulation parameters for the respective ``base'' cases. Values in parentheses indicate the respective parameters that are different for the cases with high voltage amplitude.}
    \begin{tabularx}{\textwidth}{lXX}
    \hline\hline
    Parameter & 10 Pa cases & 100 Pa cases \\ 
    \hline
    Voltage amplitude & $V=250~(400)$ V & $V=250 ~(400)$ V\\
    Frequency &$f=13.56$ MHz & $f=13.56$ MHz\\
    Gas pressure & $p=10$ Pa & $p=100$ Pa\\
    Gas temperature & $T_{\rm g}=300$ K & $T_{\rm g}=300$ K\\
    Secondary electron emission yield & $\gamma=0.1$ & $\gamma=0.1$\\
    Electron reflection coefficient & $R=0$ & $R=0$\\
    Electrode gap & $L=25$ mm & $L=25$ mm\\
    Number of grid points & $N_x=256$ & $N_x=512$\\
    Number of timesteps & $N_t=9400$ & $N_t=18800$\\
    Particle weight & $W=10^5$ & $W=5\cdot10^5$ ($3\cdot10^6$)\\
    \hline\hline
    \end{tabularx}
    \label{tab:table2}
\end{table*}

Table~\ref{tab:table2} shows the parameters of the four ``base'' cases. The initial numerical parameters are chosen to fulfill the stability criteria listed above with threshold values representing the compromise between accuracy and runtime. The voltage amplitudes were chosen to be $\phi_0=250$~V and $\phi_0=400$~V, whereas the pressures are $p=10$~Pa and $p=100$~Pa. The frequency of the excitation waveform is kept at a constant value of 13.56 MHz throughout, along with the electrode gap ($L=25$~mm) and the gas temperature ($T_{\rm g}=300$~K). The reason for not changing these parameters was to investigate what effect the numerical parameters have on the results of the simulations: ideally, changing any of these numerical parameters should not affect the results, as the same physics is described. 

In all base cases, we assume an electrode surface having an electron reflection coefficient ($R$) of 0 and a ion induced secondary electron emission coefficient ($\gamma$) of 0.1. These are the only parameters, which will also be varied but are -- strictly speaking -- not numerical parameters. The numerical parameters investigated here will be (i) the number of timesteps, $N_t$, which determines the number of times superparticles are moved in one RF-cycle. In other words, it introduces a minimal time scale, $\Delta t$, which the simulation can resolve. (ii) the number of grid points, $N_x$ gives a measure of the spatial resolution in the simulation: it defines a minimal length scale, $\Delta x$, which can be resolved by the simulation. (iii) another important quantity in table \ref{tab:table2} is the particle weight, $W$: it is defined as the number of real particles each superparticle (which are traced in the simulation) represents. The different values for the weight factor at $p=100$~Pa were chosen considering the fact, that for the weight of $W=5\cdot10^5$ for the lower voltage amplitude case the simulation did not achieve convergence. As shown in a later section, the physically more relevant quantity is $N_{\rm D}$, the number of superparticles in one Debye-sphere (or, in our case, one Debye-length), which is trivially related to $W$. In the simulation the electrodes are assumed to be infinite parallel planes, where no edge effect has to be taken into account, however, an arbitrary nominal surface area of 1~cm$^2$ is used simply to provide scaling between particle numbers and densities.

The plasma density is one of the fundamental and most useful physical quantities the PIC/MCC simulation can compute, therefore, in the following we consider this quantity, accompanied by the bulk electron energy probability function, to investigate the effect of varying different numerical parameters. 

\subsection{Effect of the timestep}

\begin{figure*}[tb]
    \centering
    \includegraphics[width=.8\textwidth]{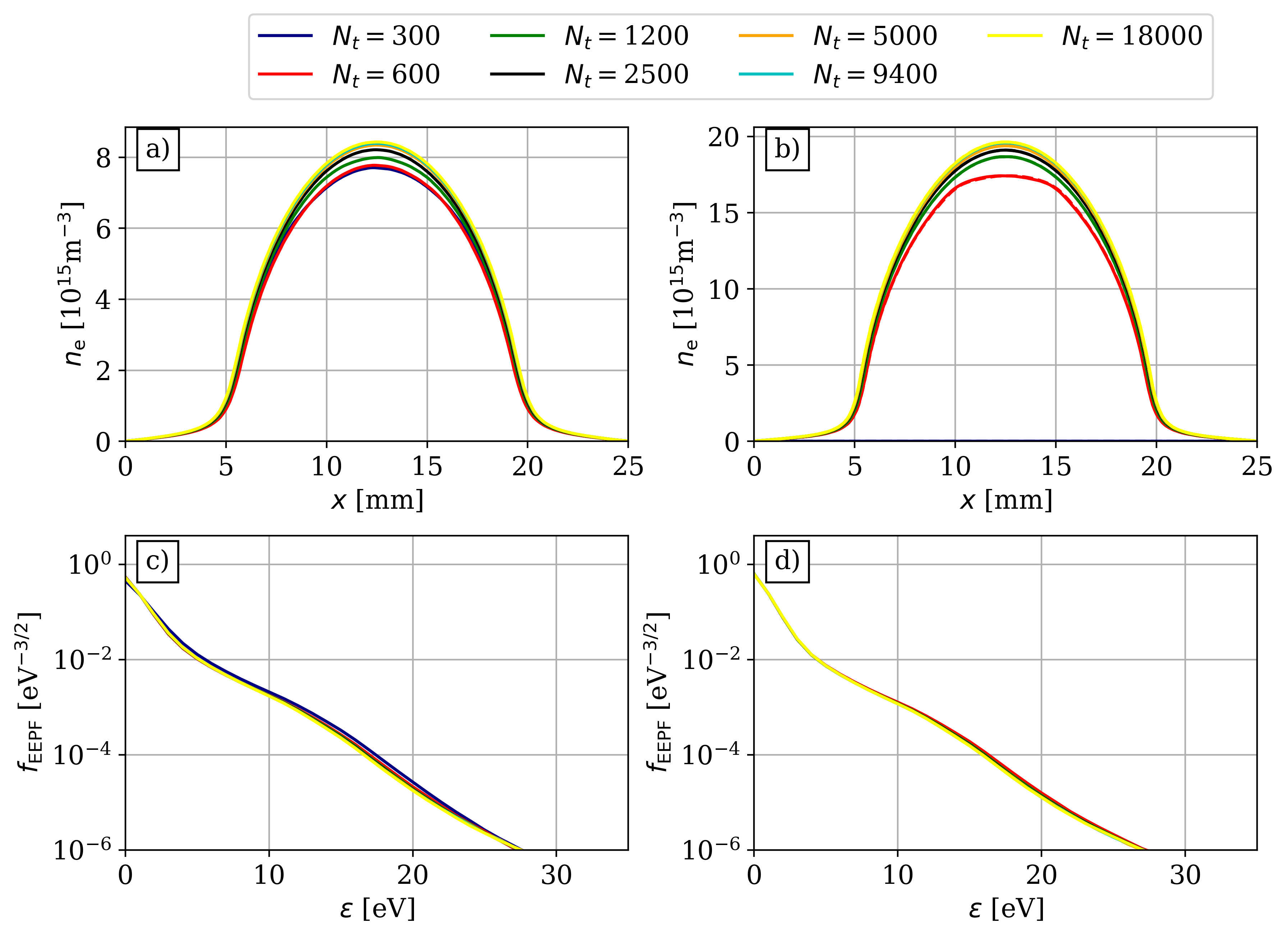}
    \caption{Time-averaged electron density distribution, $n_{\rm e}(x)$, and electron energy probability function, $f_{\rm EEPF}(\varepsilon)$, in the middle of the discharge (at $x=L/2$) for different timestep values ($N_t$) at $p=10$~Pa for $\phi_0=250$~V (a/c) and $\phi_0=400$~V (b/d). In panels (b/d) the line for $N_t=300$ is absent as this case proved to be numerically unstable.}
    \label{fig:10PT}
\end{figure*}

Figure~\ref{fig:10PT} shows the time averaged electron density, $n_{\rm e}$, and the electron energy probability function, $f_{\rm EEPF}$ at $p=10$~Pa for a voltage amplitude of $\phi_0=250$~V (a/c) and 400~V (b/d), for different values of the timestep number ($N_t$). The energy probability functions shown in this article are normalized according to $\int_0^\infty \sqrt{\varepsilon}f_{\rm EEPF}(\varepsilon)\,{\rm d}\varepsilon = 1$. As seen in panel (a), for the lower voltage amplitude, the density maxima for the two limiting cases, i.e. $N_t=300$ and $N_t=18000$, differ by less than 10\% only. By increasing the number of timesteps, the density increases monotonically, and reaches a saturation value for a high enough timestep number (here $N_t\geq5000$). The reason for the increase of the density as a function of the timestep at low pressure can be understood as follows. Having a relatively low value for the timestep introduces two effects due to the lower resolution, and, consequently, the higher $\Delta t$: (i) the trajectories of the individual superparticles will be smoothed, in the sense that for a lower $N_t$, all changes in the trajectory, that happen under the timescale of the given $\Delta t$, will be missed. This is primarily connected to the solver of the equation of motion. Using a too low $N_t$ might lead not only to inaccuracies, but to an instability as well, as in case of $\phi_0=400$~V and $N_t=300$. (ii) Similarly, by lowering the resolution of the time, potential collisions might be missed. The reason for this is, that within each timestep only one collision is handled for a given particle. If the collision probability $P_{\rm coll}$ is high enough that multiple collisions can occur (assuming independence between collisional events, the probability of having $n$ consecutive collisions in the same timestep is $\sim P_{\rm coll}^n$), these additional collisions will not be taken into account by the simulation. 

As seen in panel~(c), the electron energy probability function (EEPF) at the discharge center is a two-temperature distribution function \cite{Godyak90,Godyak2021}, i.e., there are a high number of electrons having an energy above the ionization threshold (15.76 eV). Thus, by increasing the number of timesteps, thereby reducing the probability of missing consecutive collisions in a given timestep, the number of ionization events in the simulation increases, together with the electron density. However, the actual value of $N_t$ does not seem to have a significant impact on the EEPF itself: they look fairly similar in all cases, although, for $N_t=18000$ we can observe a decrease of $f_{\rm EEPF}$ at energies above the ionization threshold: these electrons were the ones contributing to ionization and leading to the density increase as compared to the density profiles at a lower value of $N_t$. Increasing the voltage amplitude (panels b,d) leads to a similar trend in the physical quantities investigated: although, in this case, there is a threshold value for $N_t$, under which a numerical instability develops, and the next lowest timestep number ($N_t=600$) also shows significant deviation from the high time-resolution density profiles.
 
As it will be shown in section \ref{sec:stab}, the numerical instability most likely comes from the leapfrog time-integration scheme. For $N_t\geq1200$, the density differs only by a few percent. This can also be seen in panel (d) for the EEPF. By increasing the voltage amplitude, the mean energy of electrons is increased, and due to the low pressure, the high energy tail gets more pronounced. This is the reason why in panel (d) for the EEPF the two ``temperatures'' are visibly different. 

\begin{figure*}[tb]
    \centering
    \includegraphics[width=.8\textwidth]{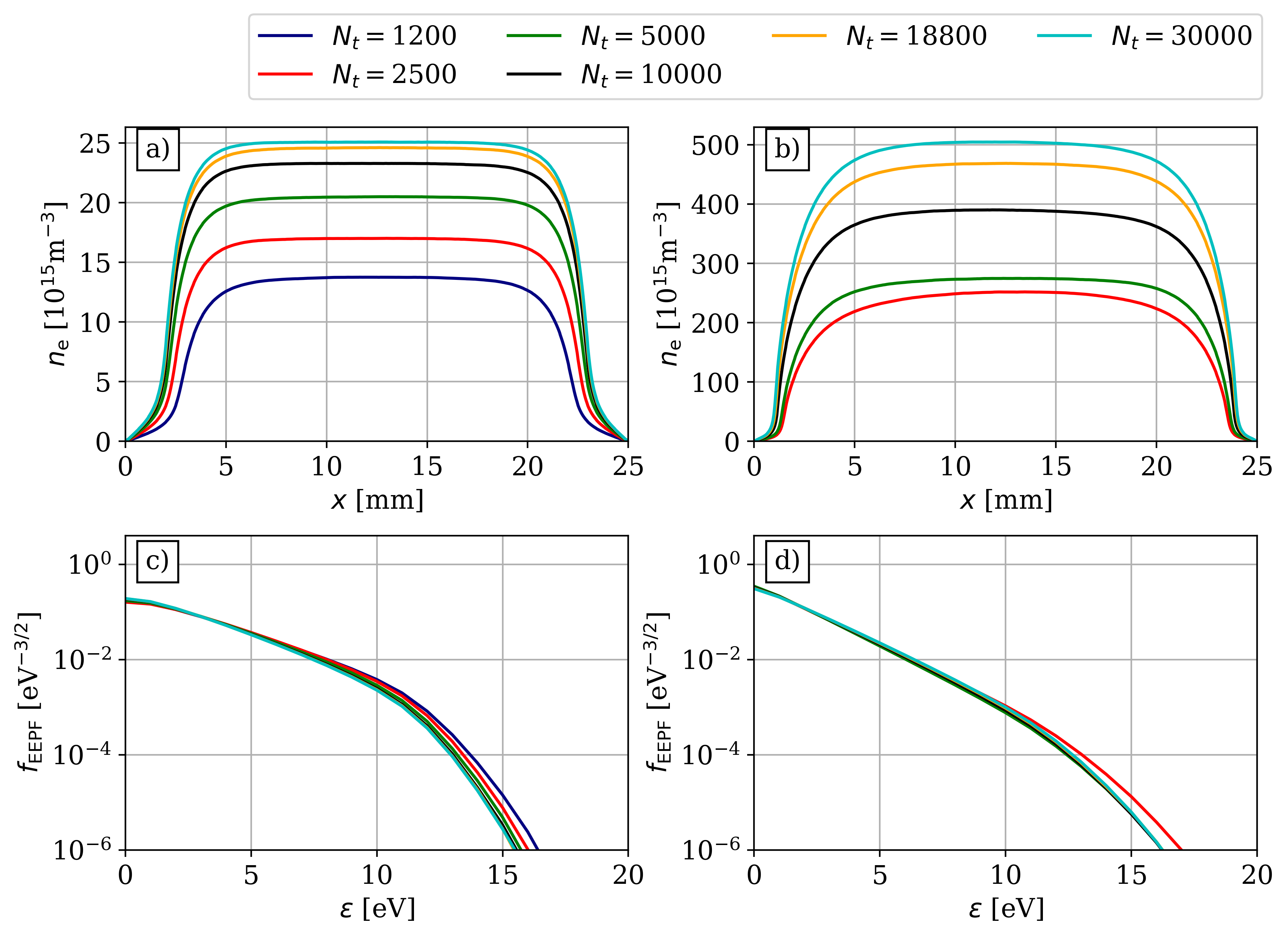}
    \caption{Time-averaged electron density distribution, $n_{\rm e}(x)$, and electron energy probability function, $f_{\rm EEPF}(\varepsilon)$, in the middle of the discharge (at $x=L/2$) for different timestep values ($N_t$) at $p=100$~Pa for $\phi_0=250$~V (a/c) and $\phi_0=400$~V (b/d). In panels (b/d) the line for $N_t=1200$ is absent as this case proved to be numerically unstable.}
    \label{fig:100PT}
\end{figure*}

Figure~\ref{fig:100PT} shows the base cases at high pressure, $p=100$~Pa. Panels (a,c) show the electron density distribution and the EEPF in the case of $\phi_0=250$~V voltage amplitude, respectively. The characteristics of the discharge at this pressure are different from the $p=10$~Pa case: the transport is more local, and ionization happens primarily near the sheath region, due to the ambipolar electric field \cite{Schulze_2014}: this is the reason for the wide plateaus in panels (a,b) for the density, and also, the Druyvesteyn-like EEPFs in the discharge center (panels c,d), which indicates the lack of ionization in this spatial region \cite{Kawamura1}. 

The difference between the densities is greater in this case, as compared to low pressure: between the lowest timestep ($N_t=1200$) and the highest ($N_t=30\,000$), the difference is almost a factor of 2. Also, at this higher pressure, a higher $N_t$ was needed to attain convergence as compared to fig.~\ref{fig:10PT}. Although the magnitudes are different, the main trend agrees with the low pressure case: $n_{\rm e}$ increases with increasing $N_t$. The reason for the amplified difference is the fact, that at this pressure the mean free path of the electrons is much smaller, and thus they undergo more collisions than at low pressure. If time does not have a fine enough resolution, many collisions, much of which, in the region of the sheaths, are electron impact ionization, will be missed, contributing to a depleted electron density. It seems, however, that for a large enough $N_t$, the density converges to a stable distribution (in panel (a) it peaks at ${\hat n}_{\rm e}\sim2.5\cdot10^{16}$ m$^{-3}$). The role of missed collisions is also evident from panel (c), the shape of the EEPF: at low $N_t$, the high energy tail is lifted to higher energies: an indicator that electrons can reach high energies without undergoing collisions, some of which, for a higher value of $N_t$, might cause ionization, thereby increasing the electron density.

A similar trend is observable for the high voltage amplitude case (panels b,d). Here, the $N_t=1200$ value was found to be unstable and convergence could not be achieved. Still, the density increases as a function of the number of timesteps, resulting in a difference of more than a factor of 2 in the density between the two limiting cases. It is also worth noting, that the spatial region of the plateau, i.e. the width of the constant density region is shrunk, a clear indication of the increased sheath widths and a wider spatial region for ionization as compared to the $\phi_0=250$~V case. The corresponding EEPFs in panel (d) show higher electron temperatures (an obvious consequence of the increased voltage amplitude), but apart from the lowest $N_t$ stable case ($N_t=2500$), the shapes of the EEPFs do not show any visible difference when changing the number of timesteps.

\subsection{Effect of the spatial grid resolution}

\begin{figure*}[tb]
    \centering
    \includegraphics[width=.8\textwidth]{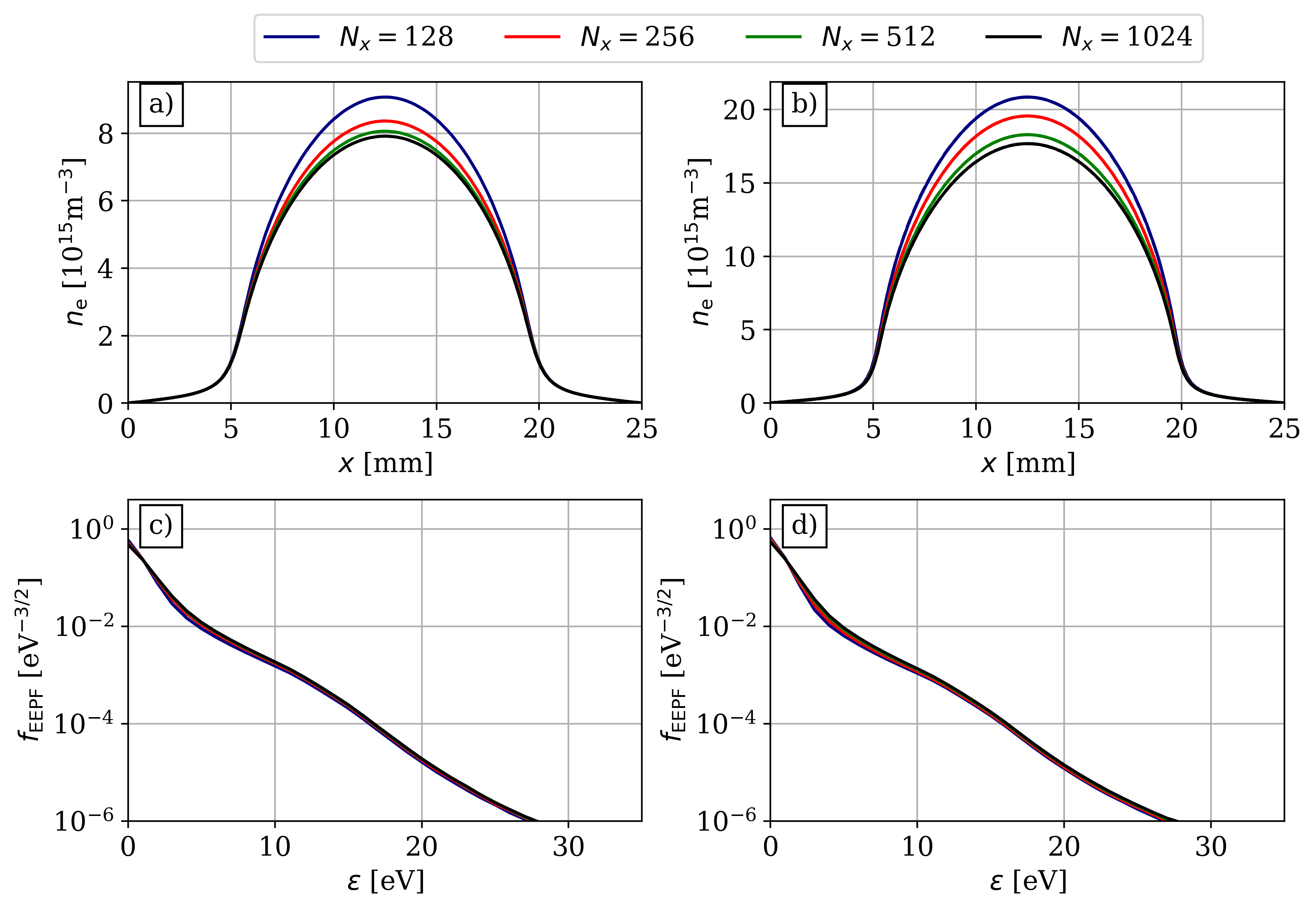}
    \caption{Time-averaged electron density distribution, $n_{\rm e}(x)$, and electron energy probability function, $f_{\rm EEPF}(\varepsilon)$, in the middle of the discharge (at $x=L/2$) for different numbers of grid points ($N_x$) at $p=10$~Pa for $\phi_0=250$~V (a/c) and $\phi_0=400$~V (b/d). }
    \label{fig:10PX}
\end{figure*}

As a next step, the effect of the number of grid points, $N_x$, is investigated. Figure \ref{fig:10PX} shows the time averaged electron density distribution and EEPF in case of $p=10$~Pa for voltage amplitudes of $\phi_0=250$~V (a,c) and $\phi_0=400$~V (b,d), respectively for different values of the number of grid points.

As seen in panels (a) and (b), in contrast to the timestep, when the number of grid points is increased, the electron density decreases, albeit the difference in density between the two limiting cases is within 10~\%. A similar trend is observable for the EEPFs for both voltage amplitudes (panels (c,d)). As the number of grid points is increased, the number of electrons having an energy higher than $\varepsilon \approx2$~eV is increased. These observations can be explained based on the fact that the number of grid points is crucial for the accurate solution of Poisson's equation and thus the correct resolution of the spatio-temporal electric field. From the perspective of electron power absorption, the main energy coupling process, the narrow region between the instantaneous sheath edge and the position of maximum sheath extent is most relevant. At positions closer to the electrode, inside the sheath, the electric field is much larger, however, the electron density is very low, while inside the bulk the electron density has its maximum but, at the same time, the electric field is negligible, both cases resulting in low power absorption. The electric field, in this most relevant region, decelerates electrons arriving from the bulk and accelerates electrons coming from the sheath towards the bulk. It consists of the contribution of the inner edge of the sheath electric field and the ambipolar electric field of the pre-sheath~\cite{Schulze_2014}. Accurate simulations need to compute and resolve this electric field structure. In the low pressure case, as the transport is non-local, and since the mean free path is large with respect to the discharge gap, many electrons can fly a long distance without undergoing collisions, thus the contribution of the electric field to the kinetic energy accumulates over a long time and distance. If the number of grid points is too small, the structure of the electric field in the pre-sheath is not resolved properly and the force on the electrons is not calculated precisely. {\color{black} Another important consequence of changing $N_x$, while keeping the number of superparticles fixed, is the presence of statistical fluctuations and, correspondingly, numerical heating, as will be discussed in sec. \ref{sec:W}. In the current case, as the number of total particles is kept fixed, as $N_x$ increases, the number of superparticles per grid cell decreases, which can increase the presence of statistical fluctuations and thus the accurate calculation of the charge density. However, a too low $N_x$ has also severe consequences on the results, as then the Debye-length is not resolved properly, and thus the charge density will be ``smoothed out'', resulting in a smaller energy increase for electrons, which then decreases ionization and thus the density. This can be seen in panel (b), where in case of an increased voltage amplitude the difference between the low $N_x$ cases and the increased $N_x$ case is greater, as due to the increased mean energy the Debye-length is smaller, and thus a higher $N_x$ is needed for the accurate description of the underlying physics.   }


\begin{figure*}[tb]
    \centering
    \includegraphics[width=.8\textwidth]{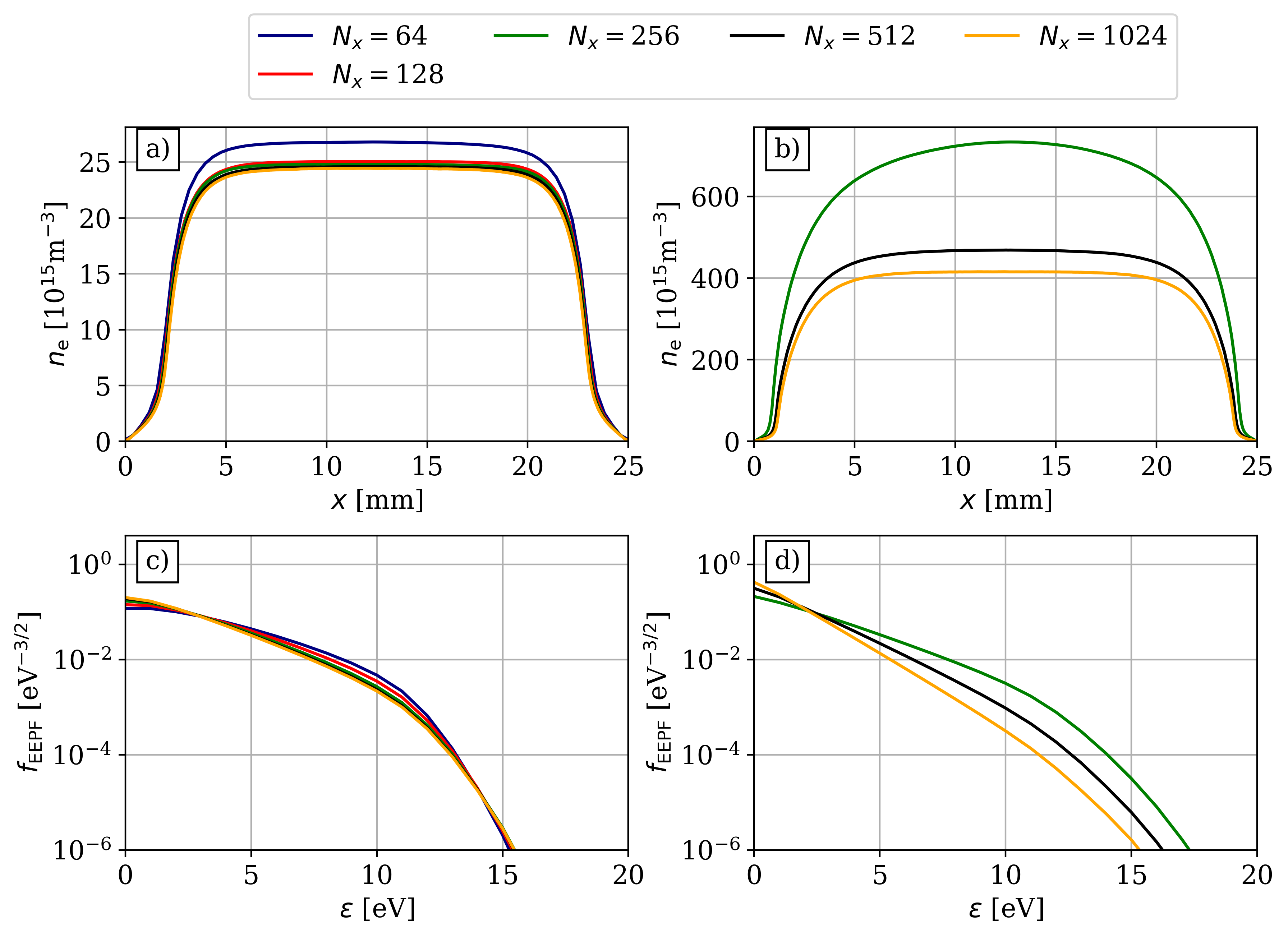}
    \caption{Time-averaged electron density distribution, $n_{\rm e}(x)$ and electron energy probability function, $f_{\rm EEPF}(\varepsilon)$, in the middle of the discharge ($x=L/2$) for different numbers of grid points ($N_x$) at $p=100$~Pa for $\phi_0=250$~V (a/c) and $\phi_0=400$~V (b/d). In panels (b/d) the lines for $N_x=64,128$ are absent as these cases proved to be numerically unstable.}
    \label{fig:100PX}
\end{figure*}

Increasing the pressure shows a different behavior: in fig.~\ref{fig:100PX}(a) the density decreases again as a function of $N_x$, but if the resolution is not too low (in this case $N_x>64$), the density stays approximately the same, with only a few percent difference between each case. In panel (c), the EEPF decreases for intermediate energies ($\varepsilon \approx 6$~eV and above), and increases for low electron energies as the number of grid points is increased. These observations can be readily explained based on the arguments presented for the low pressure case: if the pressure is increased, the transport will be more localized. This can also be seen from the Druyvesteyn-like EEPFs in the middle of the discharge with essentially zero electrons above the ionization threshold and the wide plateau of constant electron density in panel (a). This means, that there is essentially no ionization within the bulk, only near the sheath, where the sheath electric field, assisted by ambipolar field, leads to the ionization of the background gas. Thus, if the sheath electric field and the ambipolar field are resolved correctly, the density will not change much by further increasing the number of grid points. {\color{black} However, by increasing $N_x$, the statistical fluctuations are also increased. As shown in sec. \ref{sec:W}, at high pressure, the artificial heating gradually changes the Druyvesteyn-like EEPF to a Maxwellian, which has the effect to ``smooth'' out the intermediate energy regime, just as shown in panel (c).}

Nonetheless, this difference does not show up in the density itself: therefore, it can be stated, that at high pressure, when the transport is increasingly local, the number of grid points, unless the sheath electric field is poorly resolved due to the low value of $N_x$, does not lead to a further change in the electron density. The same argument cannot be used for the case of high voltage amplitude. In panel (b) we see that (i) the low $N_x$ cases were found to be unstable, thus there is a lower threshold value for the grid division as well, which must be met in order to get numerically stable results, and (ii) there are considerable differences between the electron density maxima corresponding to the different grid resolutions. {\color{black} As will be shown in sec. \ref{sec:stab}, this is mainly due to the fact, that the Debye length is not resolved properly. }One can also see in panel (d), for the EEPF, that by decreasing the number of grid points, the number of electrons, having energies higher than the ionization threshold in the discharge center, increases. Apparently, as in the high pressure and high voltage case the plasma density is the largest, resulting in very narrow electrode sheath regions, an increased spatial resolution is necessary to accurately resolve the electric field structure. On the other hand, although in 1d3v simulations the Poisson solver is computationally simple and fast, the number of grid points can not be increased without limits, at least not without increasing the number of superparticles simultaneously, as that would lead to increased numerical heating due to statistical undersampling of the charged particle density distribution, as it will be discussed below.
{\color{black}  We note, that the stability/accuracy thresholds for the grid division depends on the physical conditions and the simulation method used. In this work we only concentrate on momentum conserving PIC simulations, however, e.g., in energy conserving simulations the number of grid divisions can be much smaller without reaching an instability~\cite{Chin1,Eremin}.}

\subsection{Effect of the electrode surface model}

{\color{black} In this section the effect of the parameters describing a simple, albeit widely used electrode surface model is presented. The model involves two coefficients, the electron reflection coefficient, $R$, and the ion induced secondary electron emission coefficient (SEEC), $\gamma$. We do not aim to discuss more complex, realistic models of electrode surfaces \cite{gam1,gam2,gam3,gam4}.

Although, strictly speaking, these are not numerical parameters, like the ones considered up to now, as these describe different physical conditions, whereas the other parameters investigated should ideally describe the ``same physics'', it is nonetheless an interesting question when any of these parameters can be neglected altogether, i.e. in which of the four physical base cases investigated here can some aspect of the electrode surface be considered irrelevant.}

\begin{figure*}[htb]
    \centering
    \includegraphics[width=.8\textwidth]{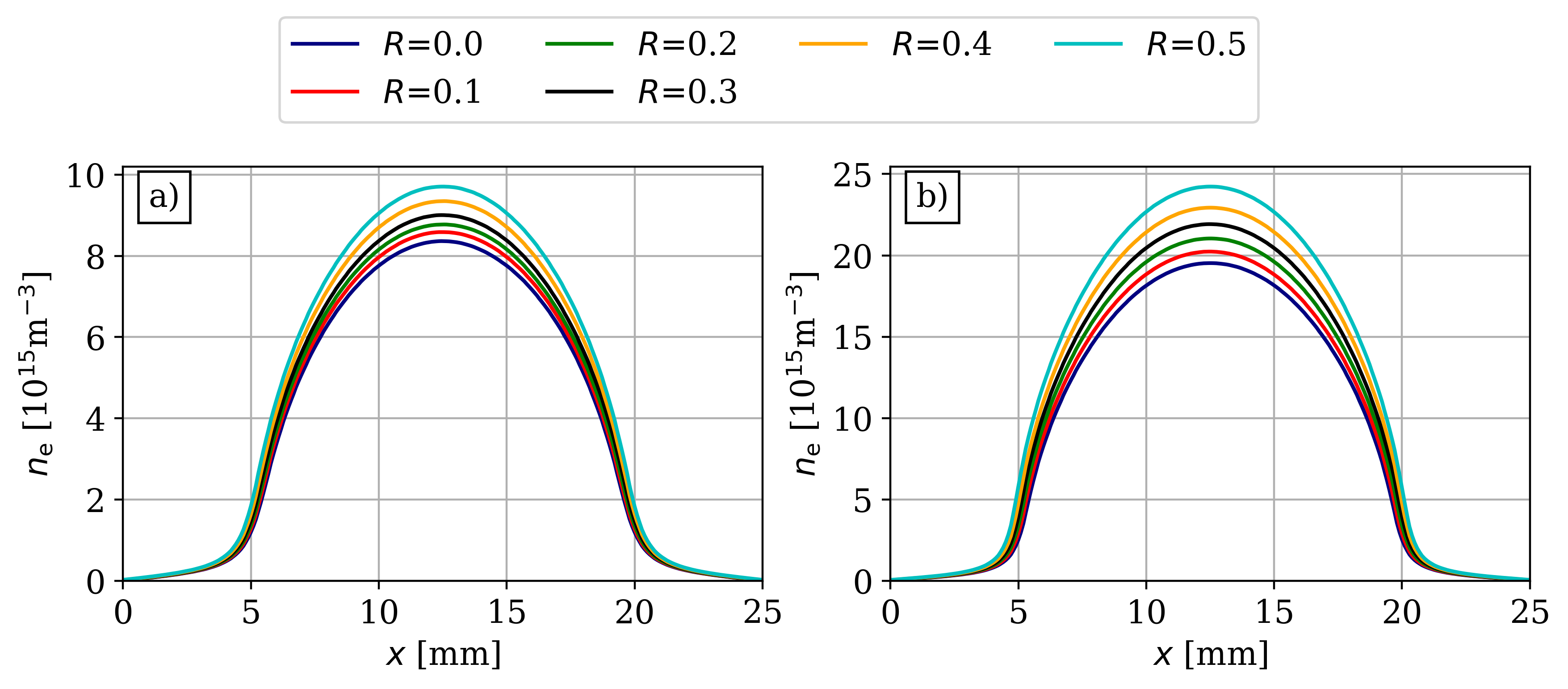}
    \caption{Time-averaged electron density distribution, $n_{\rm e}(x)$ for different electron reflection coefficients ($R$) at $p=10$~Pa for $\phi_0=250$~V (a) and $\phi_0=400$~V (b).}
    \label{fig:10PR}
\end{figure*}

Figure~\ref{fig:10PR} shows the time averaged electron densities for different electron reflection coefficients ($R$) at $p=10$~Pa for $\phi_0=250$~V (a) and 400~V (b). In both cases the electron density monotonically increases by increasing the electron reflection coefficient; the reason for this is that electrons, which were to be lost in case of $R=0$, are reflected back into the plasma and can contribute to ionization: the higher $R$ is, the more pronounced the effect can be. Similarly, by increasing the voltage amplitude from $\phi_0=250$~V to 400~V, the increase in density is more drastic (the difference between the $R=0$ and $R=0.5$ cases is $\approx10\%$ in panel (a), while $\approx20\%$ in panel (b)). In this case, due to the higher voltage amplitude, reflected electrons have a higher chance to ionize, thus increasing the electron density. Based on this the electron reflection coefficient has a non-negligible effect on the plasma density at low pressure.

\begin{figure*}[htb]
    \centering
    \includegraphics[width=.8\textwidth]{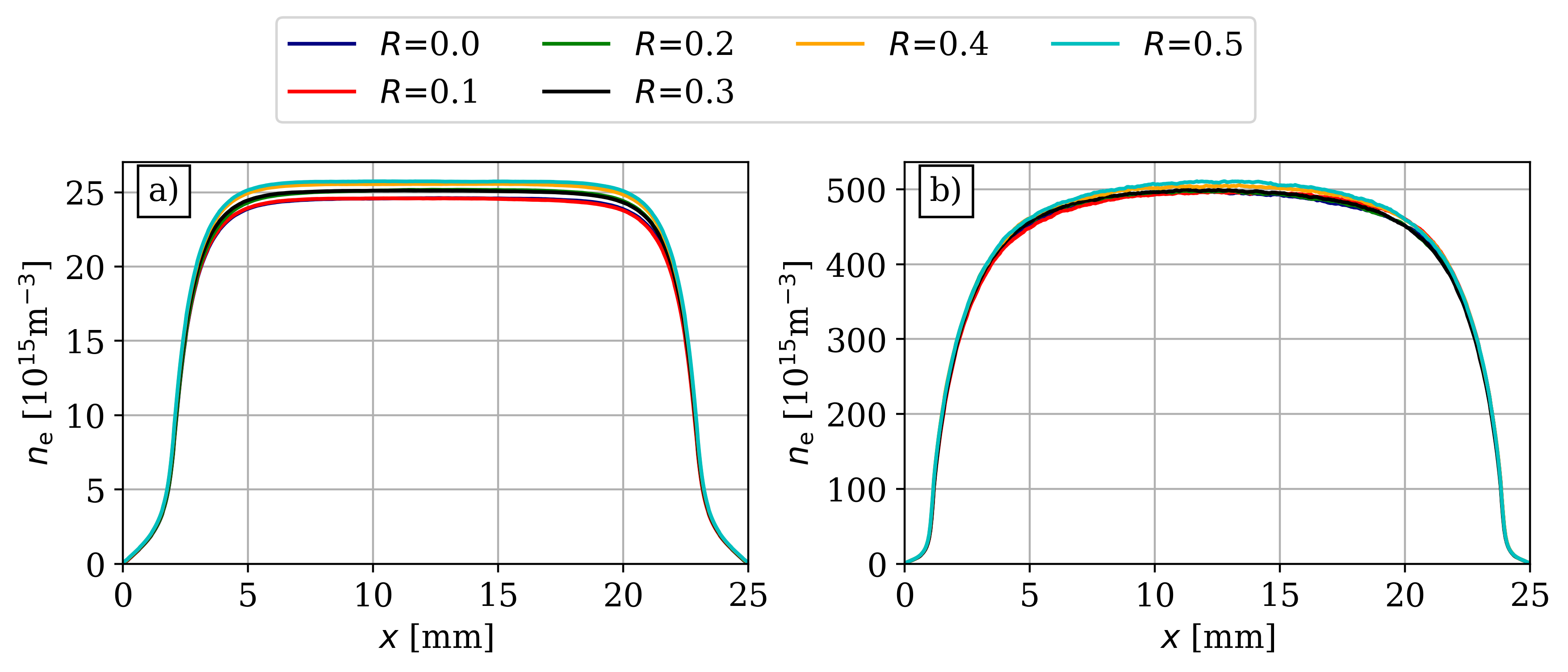}
    \caption{Time-averaged electron density distribution, $n_{\rm e}(x)$ for different electron reflection coefficients ($R$) at $p=100$~Pa for $\phi_0=250$~V (a) and $\phi_0=400$~V (b).}
    \label{fig:100PR}
\end{figure*}

Increasing the pressure from $p=10$~Pa to 100~Pa, fig.~\ref{fig:100PR} shows the mean electron densities for $\phi_0=250$~V (a) and 400~V (b) for the same reflection coefficients, as before. In contrast to the low pressure case, at high pressure the electron reflection coefficient seems to have a negligible effect on the plasma density, regardless of the voltage amplitude: the difference between the cases with the $R=0$ and $R=0.5$ values is only a few percent. 

The reason for this behaviour is the local nature of the transport: as there are few energetic electrons at this high pressure, electrons can only reach the electrode when the sheath is close to its collapsed phase. Due to the small mean free path, these electrons can seldom reach a high enough energy to ionize: they usually lose their energy in frequent elastic collisions. Thus, at high pressure the exact value of the electron reflection coefficient seems not to play an overly important role in the plasma behaviour.

\begin{figure*}[htb]
    \centering
    \includegraphics[width=.8\textwidth]{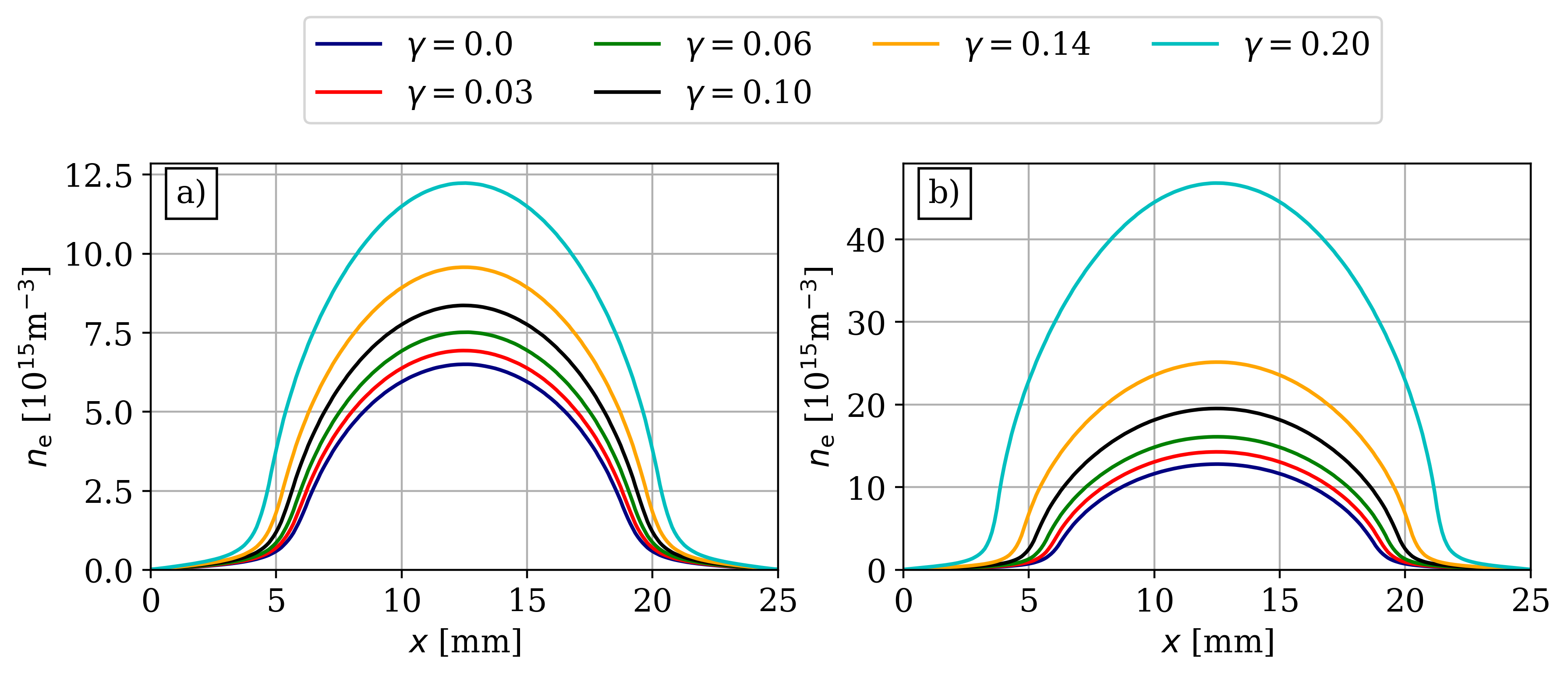}
    \caption{Time-averaged electron density distribution, $n_{\rm e}(x)$ for different SEECs ($\gamma$) at $p=10$~Pa for $\phi_0=250$~V (a) and $\phi_0=400$~V (b).}
    \label{fig:10PG}
\end{figure*}

Figure~\ref{fig:10PG} shows the time-averaged electron density at $p=10$~Pa and voltage amplitudes $\phi_0= 250$~V (a) and 400~V (b) for different secondary electron emission coefficients, $\gamma$. In case of $\phi_0=250$~V, the electron density monotonically increases as a function of $\gamma$. This is understandable, as $\gamma$ gives the probability of one ion leading to the emission of an electron when reaching one of the electrodes. As ions continuously flow towards the electrodes, secondary electrons can gain considerable energy due to the high sheath electric field, thereby contributing more to ionization than
those electrons, which undergo reflections (cf. fig.~\ref{fig:10PR}). Consequently, increasing the voltage amplitude, thereby increasing the magnitude of the sheath electric field, leads to an even sharper increase of the electron density, as seen in fig.~\ref{fig:10PG}(b). 

\begin{figure*}[htb]
    \centering
    \includegraphics[width=.8\textwidth]{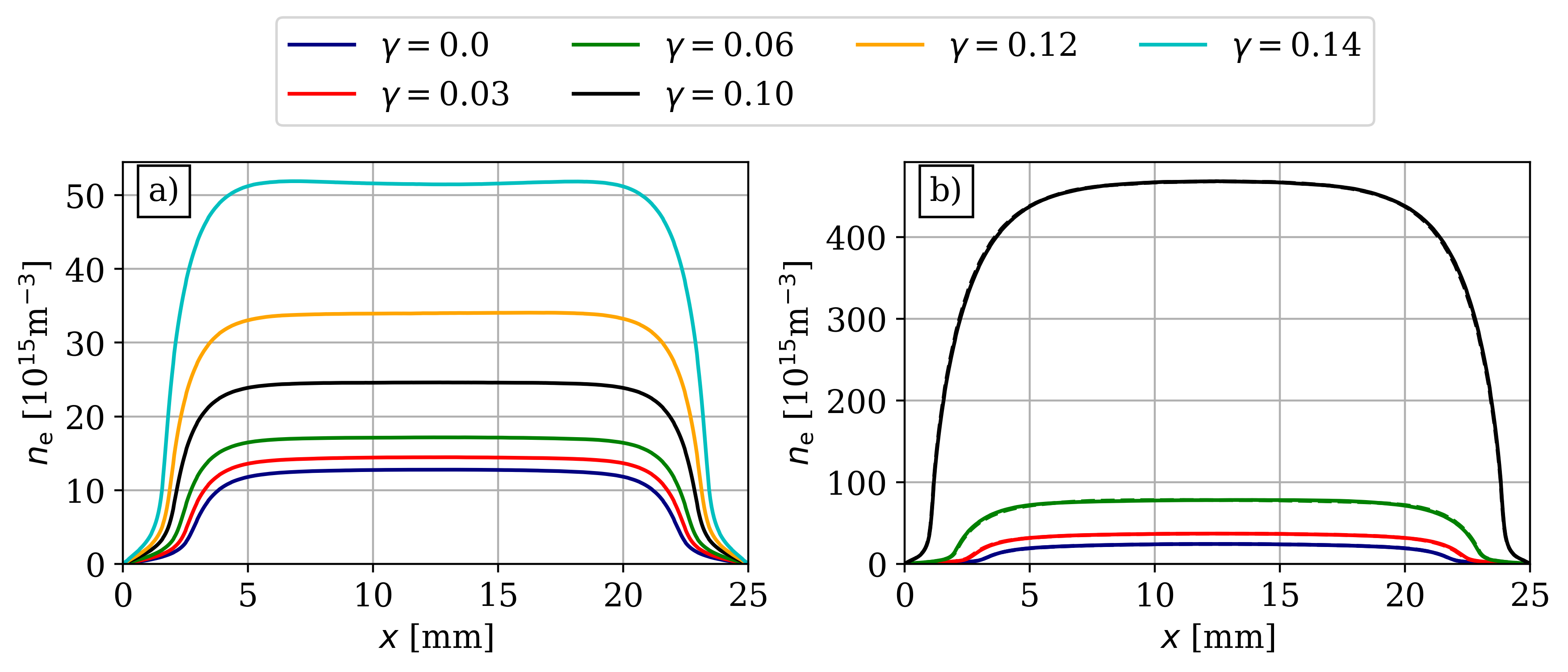}
    \caption{Time-averaged electron density distribution, $n_{\rm e}(x)$ for different SEECs ($\gamma$) at $p=100$~Pa for $\phi_0=250$~V (a) and $\phi_0=400$~V (b).}
    \label{fig:100PG}
\end{figure*}

Increasing the pressure in case of nonzero secondary electron emission, the monotonic increase of the electron density as a function of $\gamma$ persists, as seen in fig.~\ref{fig:100PG}. For $\phi=250$~V (panel (a)) the difference between the density maxima for the two extreme cases, i.e. $\gamma=0.0$ and $\gamma=0.14$ is a factor of $\sim4$. For the high amplitude case (i.e. $\phi_0=400$~V) the plasma proved to be unstable for $\gamma>0.1$, which is an indication of arcing, which the simulation cannot handle. Still, it can be seen, that there is a density increase of a factor of 4 between $\gamma=0.06$ and $\gamma=0.1$, an indication of exponential increase of the density as a function of the secondary electron emission coefficient. Thus, overall, it can be stated, that the secondary electron emission coefficient, as opposed to the electron reflection coefficient, is important at high pressure as well, and should be taken into account when modelling the electrode surfaces. 

\subsection{Effect of the particle weight}\label{sec:W}

Arguably, one of the most interesting numerical parameters in PIC/MCC simulations is the superparticle weight, $W$, or, equivalently, the number of superparticles considered in the simulation. As PIC/MCC simulations use a mean-field like approach, its validity assumes a high number of particles within one Debye-sphere, i.e. $N_D\gg1$ (see Sec.~\ref{sec:intro}) \cite{turner2006kinetic}. Although this constraint is not entirely well defined, one would assume that the higher the number of particles (or, in our case, the lower the weight) is, the more accurate description the PIC/MCC method can provide. 

\begin{figure*}[htb]
    \centering
    \includegraphics[width=.8\textwidth]{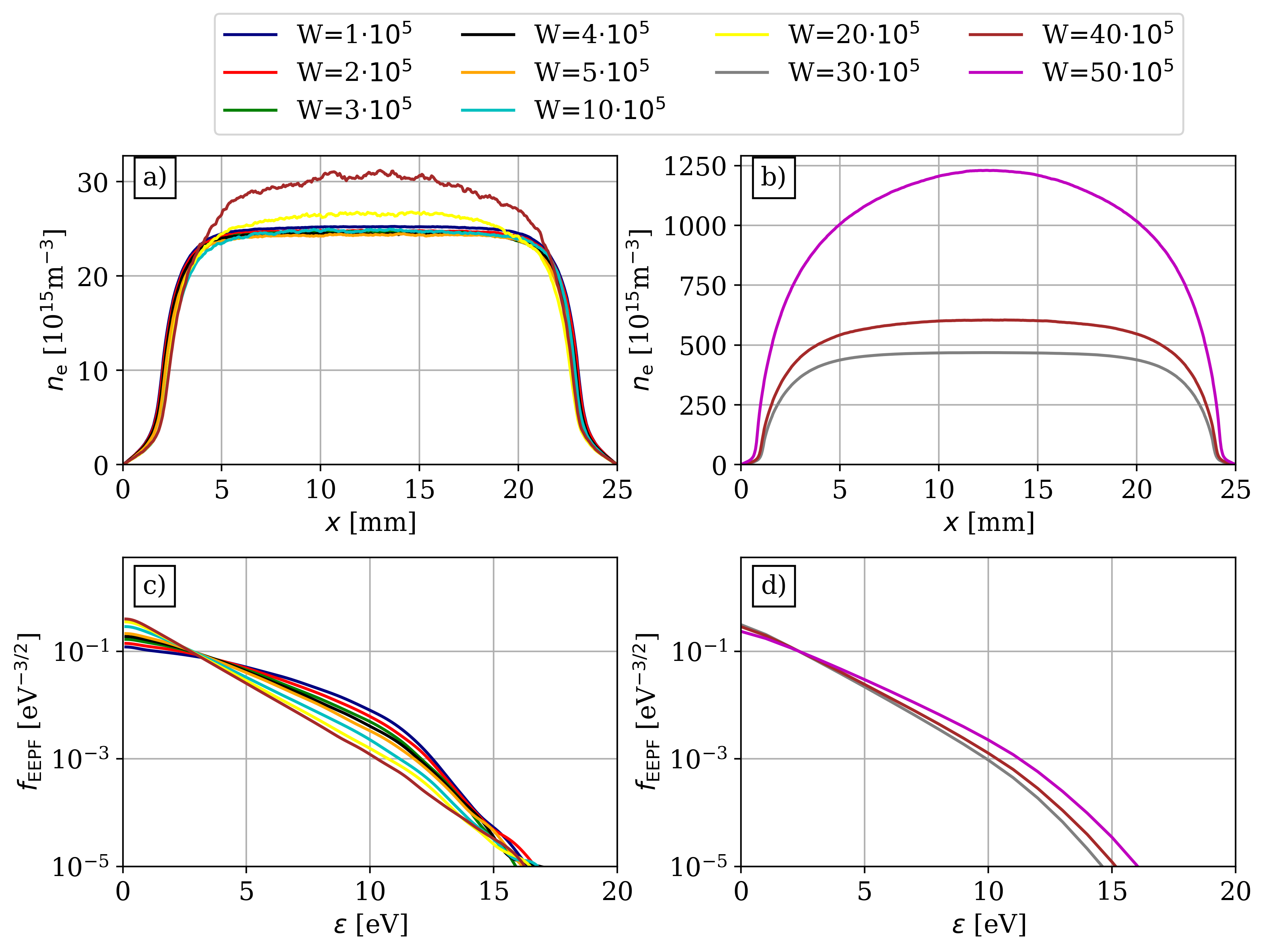}
    \caption{Time-averaged electron density distribution, $n_{\rm e}(x)$ and electron energy probability function, $f_{\rm EEPF}(\varepsilon)$, in the middle of the discharge ($x=L/2$) for different values of the particle weight ($W$) at $p=100$~Pa for $\phi_0=250$~V (a/c) and $\phi_0=400$~V (b/d).}
    \label{fig:100PW}
\end{figure*}

To investigate the validity and consequences of this argument, for reasons that will become obvious below, we first introduce the high pressure results. Figures~\ref{fig:100PW}(a) and (c) show the electron density distribution and EEPFs in case of $\phi_0=250$~V voltage amplitude, respectively, at a pressure of $p=100$~Pa. Interestingly, increasing the number of particles, at least initially, leads to a decrease of the electron density. For the EEPFs (panel (c)) a transition from a Maxwellian at low particle number (high particle weight) to a Druyvesteyn-like EEPF at high particle number can be observed. This partially explains the decrease of the density: there is more ionization for a Maxwellian distribution than for a Druyvesteyn-like EEPF (assuming identical mean energies in both cases). This ``ensemble description'' can be complemented by a ``particle description'', if we consider the effect of individual fluctuations on the local transport at high pressure. 

The ``artificial Maxwellianization'', also referred to as numerical heating, can be understood based on the observations made above: if the statistical fluctuations of the physical quantities (e.g. electric field) get increasingly important, these ``random events'' lead to an isotropisation of the electrons' motion and thus to an EEPF close to a Maxwellian. This artificial Maxwellianization has already been reported by Turner \cite{turner2006kinetic}. At high particle numbers, where the contribution of individual particle fluctuations is reduced, the mean-field description of the PIC/MCC simulations becomes increasingly accurate. In this case no isotropisation occurs in the EEPF; it changes from a Maxwellian to a two-temperature EEPF with increasing particle number. 

As seen in panel (a), the increase of the density primarily happens in the the central region of the density plateau, where, in case of a high particle number, no ionization occurs due to the low electric field and the frequent elastic collisions (which is a result of the increased pressure). Thus, when there is a small number of particles considered by the simulation, due to the local nature of the transport, the local statistical fluctuations can lead to heating and artificially increased ionization. As a result, the density is increased in panel (a). However, increasing the number of particles soon leads to a reduction of this effect, and the electron density settles (in this case at around $W=10^6$, which corresponds to approx.~50\,000 particles). Apparently, at higher pressures the choice of the particle weight is not that crucial, as the density stays constant below an upper threshold value of $W$, which, in our case could be achieved at a relatively low particle number. The high voltage amplitude case proved to be rather ``pathological'' when it comes to changing the particle number. As the density increases when the number of particles is decreased, below a certain value of $W$ the simulation was found not to converge. Thus, even for the particle weight, an instability condition can be found. On the other hand, when the number of particles is increased, due to the high number of collisions, the simulation time quickly rose. Still, the same trend can be observed here as for the $\phi_0=250$~V case: increasing the number of particles leads to the decrease of the density, and an EEPF that is more Druyvesteyn-like.

\begin{figure*}[htb]
    \centering
    \includegraphics[width=.8\textwidth]{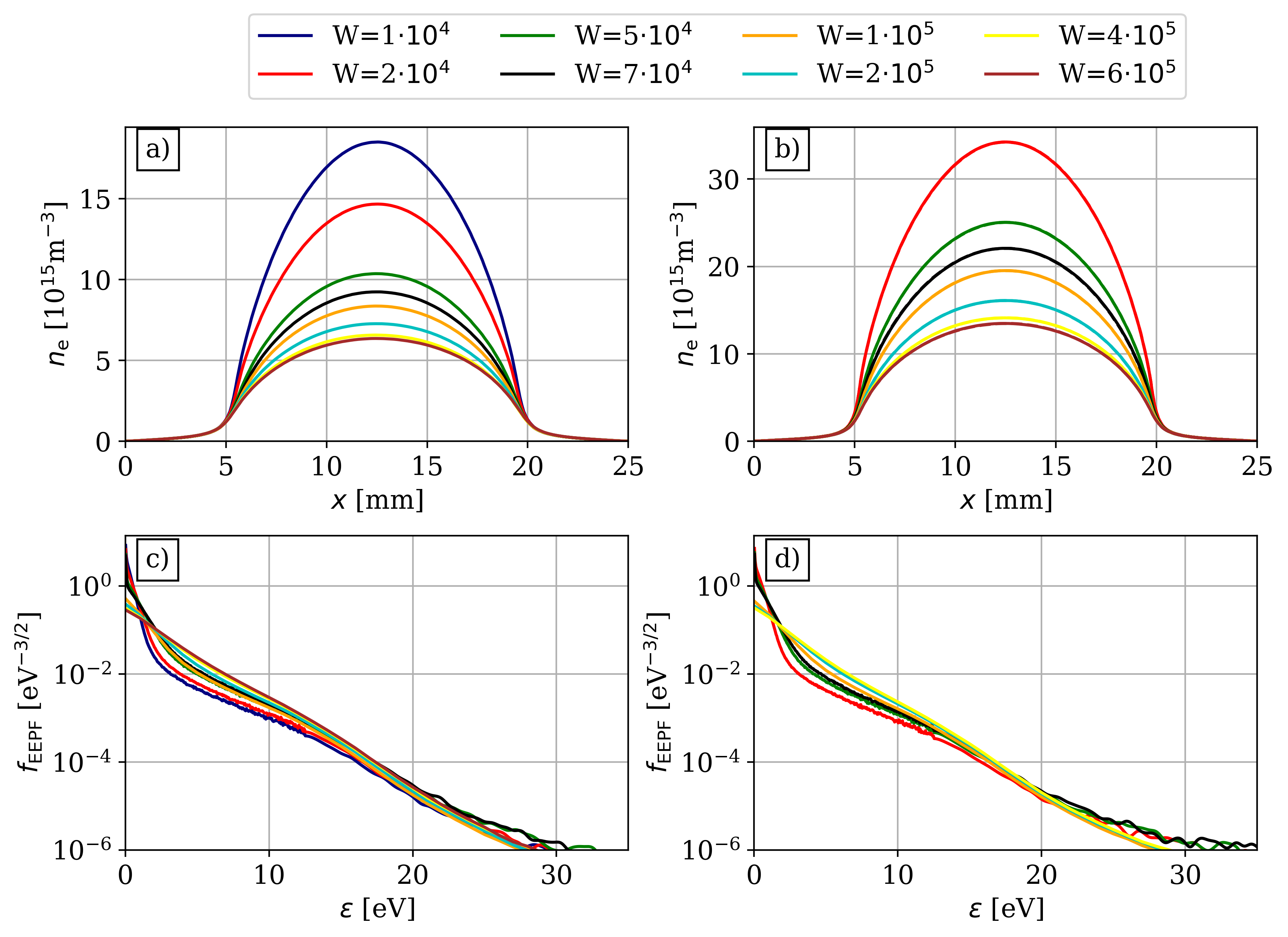}
   \caption{Time-averaged electron density distribution, $n_{\rm e}(x)$ and electron energy probability function, $f_{\rm EEPF}(\varepsilon)$, in the middle of the discharge ($x=L/2$) for different values of the particle weight ($W$) at $p=10$~Pa for $\phi_0=250$~V (a/c) and $\phi_0=400$~V (b/d). }
    \label{fig:10PW}
\end{figure*}

For the low pressure cases, fig.~\ref{fig:10PW} shows the mean electron density and EEPF for different values of the particle weight at $p=10$~Pa. The highest value of the weight factor, $W=6\cdot10^5$, corresponds to approx.~13\,000 and 25\,000 electron superparticles for a voltage amplitude of $\phi_0=250$~V (a,c) and 400~V (b,d), respectively, while the lowest weight, $W=10^4$, corresponds to approx.~$2\cdot10^6$ electron superparticles in both cases. As shown in panel (a), the electron density increases with a decreasing weight factor. The density profiles do not seem to converge to any asymptotic value in the range of weight factors investigated here, and by utilizing our recent GPU accelerated simulation~\cite{Juhasz} this trend is confirmed to continue at least up to $8\cdot10^7$ superparticles in the simulation. The density ratio of the cases of lowest and highest superparticle numbers is more than a factor of 2. Thus, the simulations are very sensitive to this particular numerical parameter. 

The fact, that incorporating higher and higher particle counts in the numerical scheme, in conflict with the general assumption, does not result in a saturation of the plasma density seems to be a major issue and certainly requires deeper investigations. From the data presented in fig.~\ref{fig:10PW}, at this point, we can draw the following conclusions. The plasma density is independent from the choice of the superparticle weight factor in the sheath region of the discharge, however, shows significant variation in the bulk. The major difference between the sheath and the bulk region is the magnitude of the electric field. In geometrically symmetric, single frequency CCPs, in the sheath region the electric field oscillates between low and high values but with constant direction, always accelerating the electrons towards the bulk, and positive ions to the electrodes. In the bulk region, at such low pressure conditions, the electric field is negligible. In this, electric field free region, low energy electrons can get trapped, as already the low magnitude field present in the pre-sheath region exerts an impenetrable potential barrier for those. This process is further enhanced by the particular shape of the collision cross section of the electron-Ar system, which features a Ramsauer-Townsend minimum~\cite{Christophorou77}, which reduces the interaction with the thermal background gas. The evolution of the EEPFs supports the above described mechanism, as the increase of the low energy population is clearly visible, electrons with energies below $\varepsilon \approx 3$~eV accumulate in the bulk, reducing the mean electron energy considerably~\cite{Kaganovich92,Kaganovich98}. Due to the proper normalization of the EEPF data in panels (c) and (d), the intermediate energy electron population appears to be decreased, but this is valid in relative terms only. Apparently the ionization, driven by high energy electrons, the source of the charged particles, is largely unaffected by the choice of the weight factor, as its reaction rate variation is found to be within an approx.~10~\% range for the values of $W$ covered here. On the other hand, the trapping of low energy electrons in the bulk is balanced by the numerical heating, which is effectively reduced with increasing superparticle numbers. 

This ``bunching'' of low energy electrons could be saturated by including electron-electron repulsion in the form of Coulomb-collisions~\cite{Kaganovich92,Kaganovich98,turner2006kinetic}, which are, however, not taken into account in our simulations and are usually not considered important in low ionization degree, low temperature plasma simulations~\cite{Birdsall1997}. According to our hypothesis, Coulomb-collisions might have a stabilizing effect on the monotonic increase of the density by introducing a negative feedback mechanism through an increasing electrostatic pressure on the electrons with increasing density. 

At the high pressure conditions, as already discussed above, no divergence of the plasma density is observed with decreasing $W$. The plasma parameters converge at already fairly low particle numbers. The reason for this qualitative difference with respect to the low pressure cases is that already at $p=100$~Pa the bulk is not completely electric field free. Due to the increased collisionality of the system, the electrical conductivity of the plasma bulk is effectively reduced, resulting in the development of an alternating electric field, on the order of a few V/cm, to ensure the conduction of the electrical current~\cite{Bulk22}. This small electric field, however, is enough to compensate for the lack of electron-electron pairwise interactions and stabilize the bulk plasma density.

Returning to the low pressure conditions, the fact that the plasma density, one of the most fundamental plasma parameters, does not converge to an asymptotic value with increasing number of simulation superparticles seems to be a catastrophic failure of the model. Indeed, if the density or the low energy part of the EEDF is of primary interest, the computations are not reliable. However, typical applications that require low pressure conditions, like etching or doping, utilize the flux of high energy ions that are accelerated by the sheath electric field without the energy dispersing effect of ion-neutral collisions. The question remains, how does this weakness of the model influence the charged particle fluxes and energy distributions at the electrodes, as these are the most relevant quantities for the applications.

\begin{figure}[htb]
    \centering
    \includegraphics[width=.9\columnwidth]{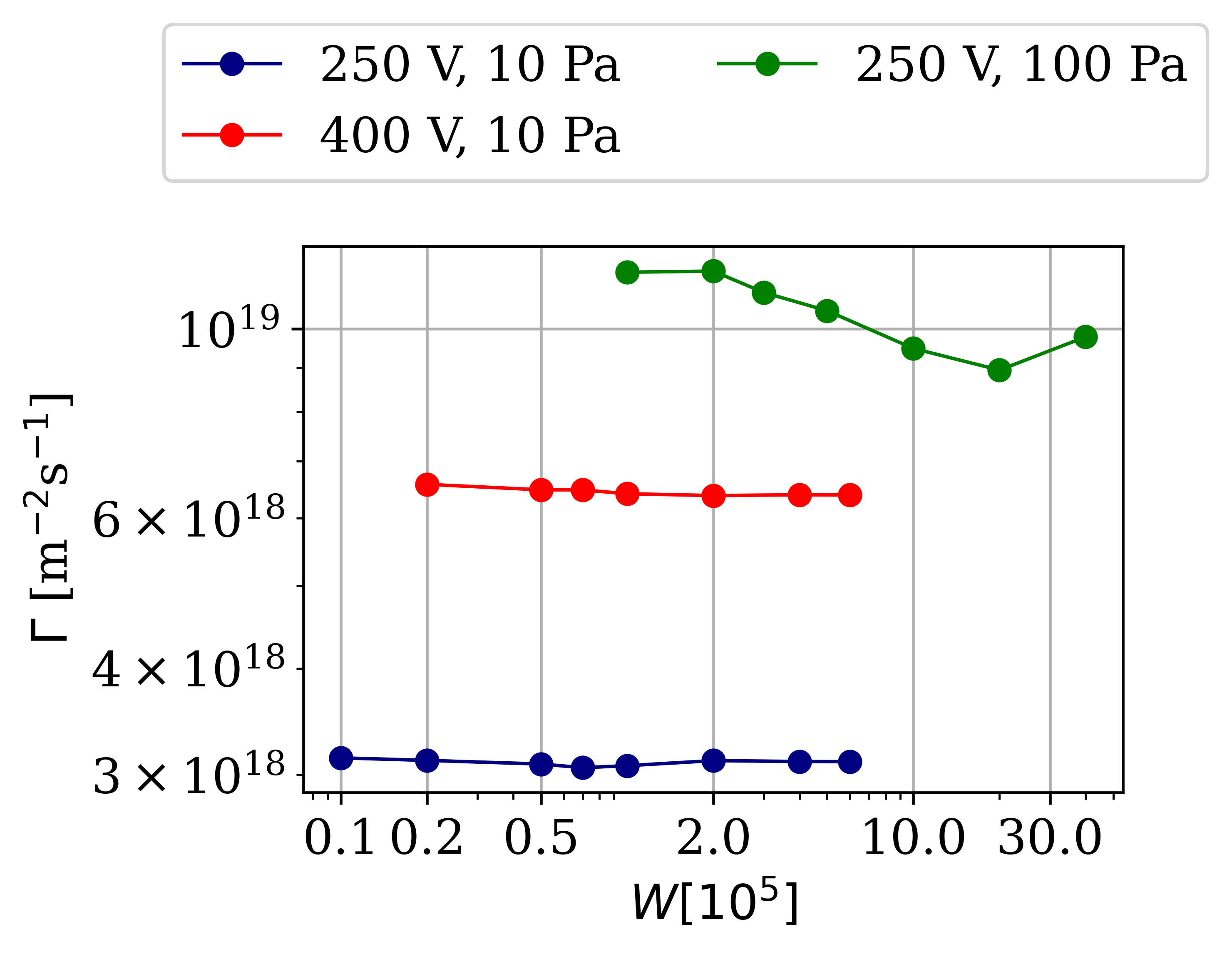}
    \caption{Mean ion flux, $\Gamma$, as a function of the superparticle weight, $W$.}
    \label{fig:Gamma} 
\end{figure}

Figure~\ref{fig:Gamma} shows the mean flux of ion arriving at the electrodes, $\Gamma$, as a function of the superparticle weight for the different cases considered. For the low pressure cases, at $p=10$~Pa, this quantity only changes within a range of $\approx 10\%$, at both low and high voltages, which is very low compared to the drastic changes of the bulk plasma density. This is consistent with the small observed variation of the ionization rate, as mentioned above, and the continuity principle stating that the sources (ionization) and losses (absorption at the electrodes) of charged particles have to be in balance in the stationary state. Similarly, for the $\phi_0=250$~V, $p=100$~Pa case (the $\phi_0=400$~V case is not shown as it became unstable), the ion flux is also within $\approx10\%$ for the four cases where the densities are converged (cf. fig.~\ref{fig:100PW}). There are deviations at lower particle numbers, but those can be attributed to the relatively large fluctuations of the electric field, which also affect the movement of the charged particles and ultimately the electron impact ionization yields.

\begin{figure*}[htb]
    \centering
    \includegraphics[width=.8\textwidth]{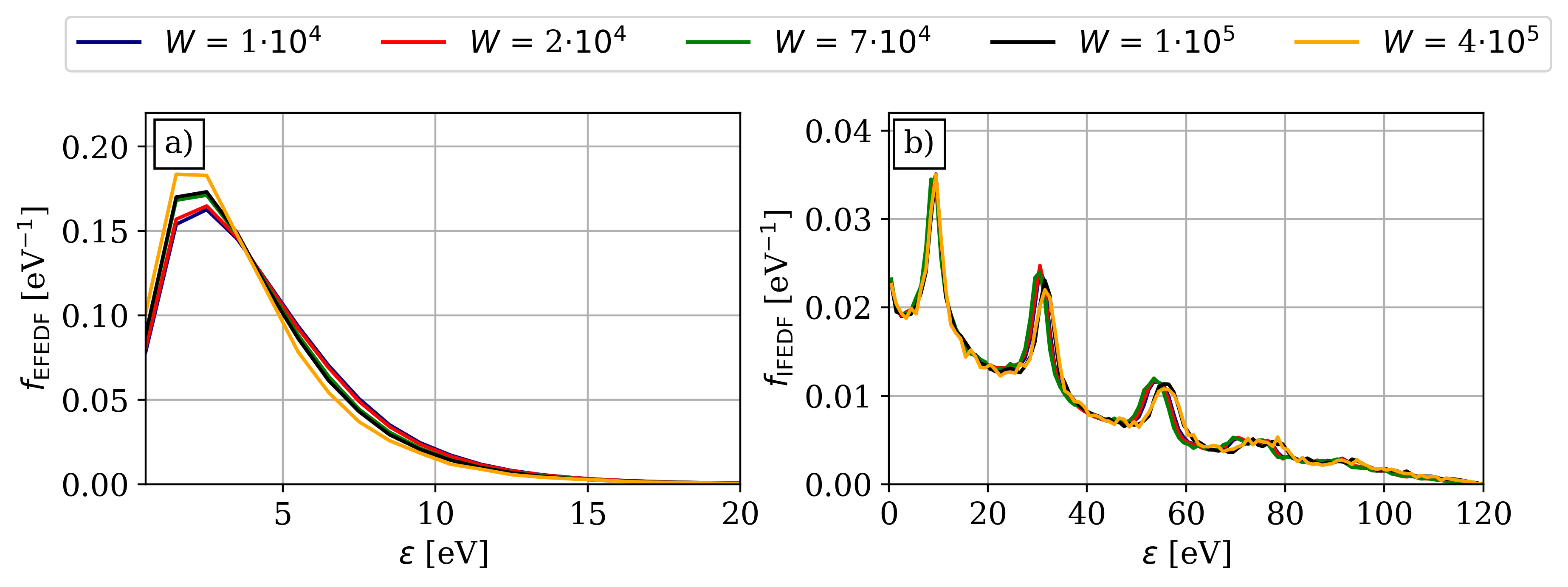}
    \caption{Time-averaged flux-energy distribution functions for (a): electrons, $f_{\rm EFEDF}(\varepsilon)$, and (b): Ar$^+$ ions, $f_{\rm IFEDF}(\varepsilon)$, at the electrodes for different values of the particle weight ($W$) at $p=10$~Pa and $\phi_0=250$~V.}
    \label{fig:FDF}
\end{figure*}

To further illustrate the stability of the application relevant plasma -- surface conditions, fig.~\ref{fig:FDF} shows the flux-energy distribution functions for electrons in panel (a) and Ar$^+$ions in panel (b). These distributions are normalized according to $\int_0^\infty f_{\rm (E/I)FEDF}(\varepsilon)\,{\rm d}\varepsilon = 1$. Even though between the lowest and highest values of the superparticle weight factor, $W$ the number of simulation particles change by a factor of $\approx 100$ and the bulk plasma density varies by more than a factor of 2, the flux-energy distributions are remarkably invariant on the particle count. The structure of the FEDFs is typical for single frequency CCPs in the low pressure, moderately collisional regime. The EFEDF peaks at low energy and decays exponentially, while the IFEDF shows a multiple-peak structure, a combined effect of periodic acceleration of ions in the sheath, and of charge exchange collisions during the ion transit~\cite{Wild91}. In the case of the highest $W$, approx.~20\,000 superparticles are present in the system for each species, which is a fairly low number causing significant statistical fluctuations and numerical heating, resulting in a slight shift towards higher energies by approx.~2~eV. Below $W\approx10^5$, with superparticle numbers exceeding $10^5$, the IFEDF does not show any dependence on $W$.

Based on these insights, we conclude that this type of simplest implementation of the electrostatic PIC/MCC simulation has difficulties to reliably predict the plasma density in CCPs at the lowest pressure conditions. However, surface processing relevant quantities, like the fluxes and energy distributions of charged particles at the electrodes are much less sensitive to the variation of some numerical parameters and can be computed reliably.

\subsection{Stability and accuracy criteria}\label{sec:stab}

In this subsection, we revisit the common stability criteria used for PIC/MCC simulations mentioned in sec. \ref{sec:intro}, based on the cases and their parameter variations described above. As already noted in sec. \ref{sec:intro}, one of the stability criteria makes use of the notion of the Debye-length, whose classical formula includes the electron temperature, or at least, the assumption of thermal equilibrium, which, especially for low pressure plasmas, is often not a well-justified assumption, as the EEPF is not Maxwellian. Still, in the following, as we only need an order of magnitude estimation, we use the classical relations, by defining a temperature that is intimately related to the mean energy of the electrons as $\langle \varepsilon\rangle=\frac{3}{2}k_{\rm B}T$. 

\begin{figure*}[htb]
    \centering
    \includegraphics[width=.8\textwidth]{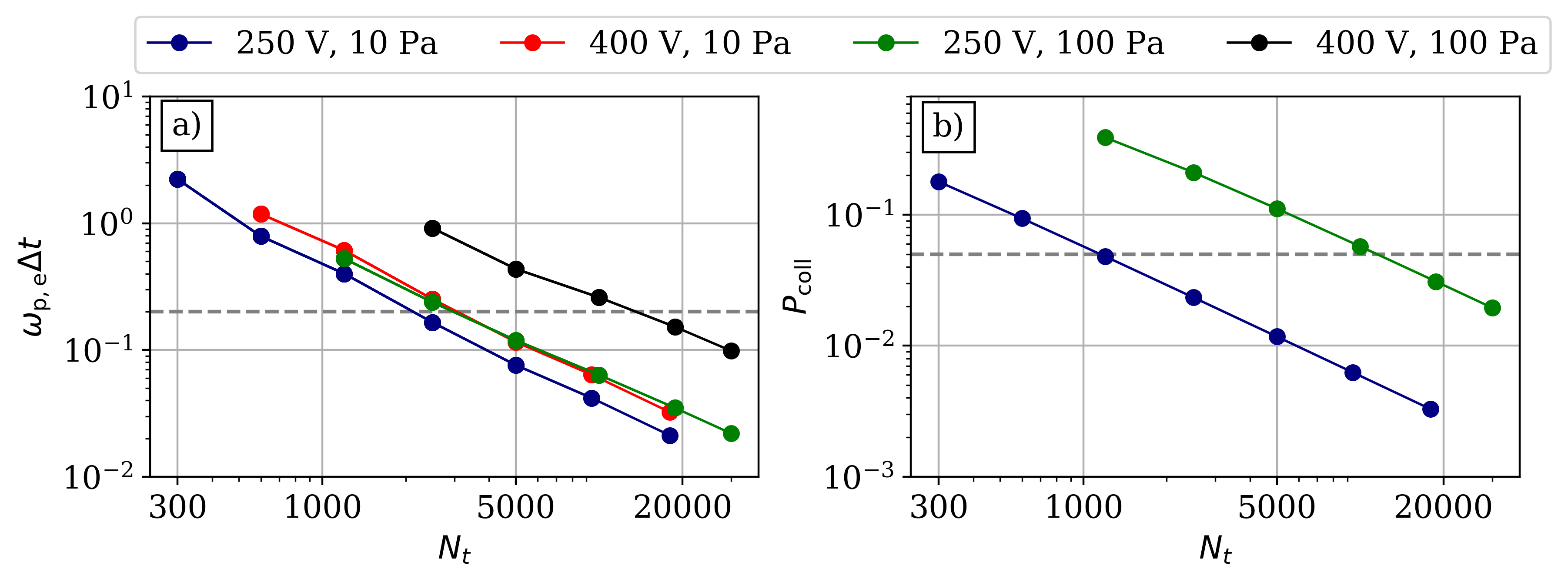}
    \caption{Ratio of the electron plasma frequency, $\omega_{\rm p,e}$, and the time sampling frequency, $\frac{N_t}{T_{\rm RF}}=\frac{1}{\Delta t}$, (a) and the collision probability of the electrons (b) as a function of the timestep, $N_t$. The dashed grey lines represent the threshold values commonly used in the literature, i.e. 0.2 for panel (a) and 0.05 for panel (b).}
    \label{fig:OmT}
\end{figure*}

Figure \ref{fig:OmT} shows the timestep-related stability criteria relevant quantities (i.e. using the plasma frequency (a) and the collision probability (b)) as a function of the number of timesteps for the different cases considered. In panel (a) the ratio of the plasma frequency, $\omega_{\rm p,e}$ and the inverse of the minimal resolved time, $1/\Delta t = N_t/T_{\rm RF}$ is shown as a function of $N_t$. The stability criterion for the leapfrog integration scheme, which is a threshold for the time difference between consecutive steps, $\Delta t$, can be derived for a harmonic oscillator with frequency $\omega_{\rm p,e}$ analytically, and is found to be $\omega_{\rm p,e}\Delta t\leq2$. \cite{Birdsall_2004} 

This only ensures numerical stability, i.e. a conserved phase space for a given electron, due to the symplectic nature of the leapfrog integration scheme. However, there can still be inaccuracies in the integration, which are on the order of $\propto(\omega_{\rm p,e}\Delta t)^2$. Thus, to ensure accurate values for the position/velocity for the particles traced, a usual threshold value of $\omega_{\rm p,e}\Delta t\leq0.2$ is chosen \cite{turner2013simulation}, which is indicated as a horizontal dashed grey line in panel (a). One can see, that although the stability condition is derived from a simple solution of the equation of motion for the harmonic oscillator, it is accurate for this complex situation as well: for $\phi_0=250$~V, $p=10$~Pa, the lowest timestep was $N_t=300$, which proved to be numerically stable, for which $\omega_{\rm p,e}\Delta t\approx2.2$. However, for $\phi_0=400$~V, $p=10$~Pa, $N_t=300$, and $\phi_0=400$~V, $p=100$~Pa, $N_t=1200$, which were numerically unstable cases, this value would be around 3. Thus, the requirement for the combination of the timestep and the inverse plasma frequency is definitely a stability condition. To answer, whether it is also an accuracy condition, we can look at the $\phi_0=250$~V, $p=10$~Pa cases, for which between the lowest and highest values of $\omega_{\rm p,e}\Delta t$ there is a difference of 3 orders of magnitude. Still, the difference in the electron density (as shown in \ref{fig:10PT} (a)) is only a few percent. Thus, although changing $\Delta t$ changes the accuracy of the solution of the equations of motion, these inaccuracies are seemingly less important e.g., the electron density.

Another very important accuracy criterion is the collision frequency, $P_{\rm coll}=1-{\rm e}^{-\nu_{\rm max}\Delta t}$, shown in panel (b). Here $\nu_{\rm max}$ is the maximum of the collision frequency, as calculated from the simulation. As the cross sections are identical for all cases, the collision probability only depends on the pressure and the timestep, but not the voltage waveform. This is why only two sets of points are shown in panel (b), together with a usual requirement of having $P_{\rm coll}\leq0.05$ \cite{turner2013simulation}, indicated as a horizontal dashed gray line. This is usually a low enough value so that multiple collisions have a very low probability to occur in a each timestep (e.g. the probability of having two collisions is less than 0.3\%). According to panel (b), in the low pressure cases, for the lowest stable timestep values the collision probabilities are $P_{\rm coll}=0.2$ and $P_{\rm coll}=0.1$ for the $\phi_0=250$~V and $\phi_0=400$~V cases, respectively. In case of $\phi_0=250$~V, $p=10$~Pa, the difference in density between the lowest and highest $N_t$ value is $\sim10\%$. Thus, the value of the collision frequency, strictly speaking, does not constitute a stability criterion, it is rather an accuracy condition, as differences in $P_{\rm coll}$ produce numerically stable, but different results. This is even more evident for the high pressure case, where electrons can only contribute to ionization in a small spatial region. If the collision probability is high, i.e. $\Delta t$ is large, some of these ionizing collisions will be missed. As shown in fig. \ref{fig:100PT}, this is indeed the case, and the differences in density can be severe, depending on the exact value of the collision frequency. However, those cases, which fulfill the $P_{\rm coll}\leq0.05$ requirement, produce densities that are close to one another. 

Now we turn to the stability and accuracy criteria which are related to the resolution of the spatial grid. There are two criteria, commonly reported in works using PIC/MCC simulations: (i) the criterion that the grid resolution, $\Delta x$ has to resolve the Debye-length, $\lambda_{\rm D}$, and (ii) the Courant-Friedrichs-Lewy (CLF) condition, which presents a constraint on both the temporal and spatial resolutions, $\Delta t$ and $\Delta x$, respectively (see sec. \ref{sec:intro}). The latter was originally used for the numerical stability of solving partial differential equations. As in our electrostatic PIC simulations only Poisson's equation is solved on the numerical grid, there is no immediate reason why this condition should be taken into account. Furthermore, the usual constraint of $v_{\rm max}\frac{\Delta t}{\Delta x}\sim1$, where $v_{\rm max}$ is the maximum velocity of the electrons, gives an overly strict constraint for stability in both the spatial and the temporal grid resolution. From our simulations we concluded that the CFL condition does not immediately apply to the electrostatic case. In many of the cases presented above there were clear violations of the CFL condition, nonetheless, were found to be numerically stable. Furthermore, the separate requirements for the spatial and temporal resolution give a more stringent threshold for the accuracy of the simulations than the CFL condition, except for some special, very low pressure and very high voltage conditions discussed elsewhere~\cite{Hartmann20,Hartmann21}.  

\begin{figure*}[htb]
    \centering
    \includegraphics[width=.8\textwidth]{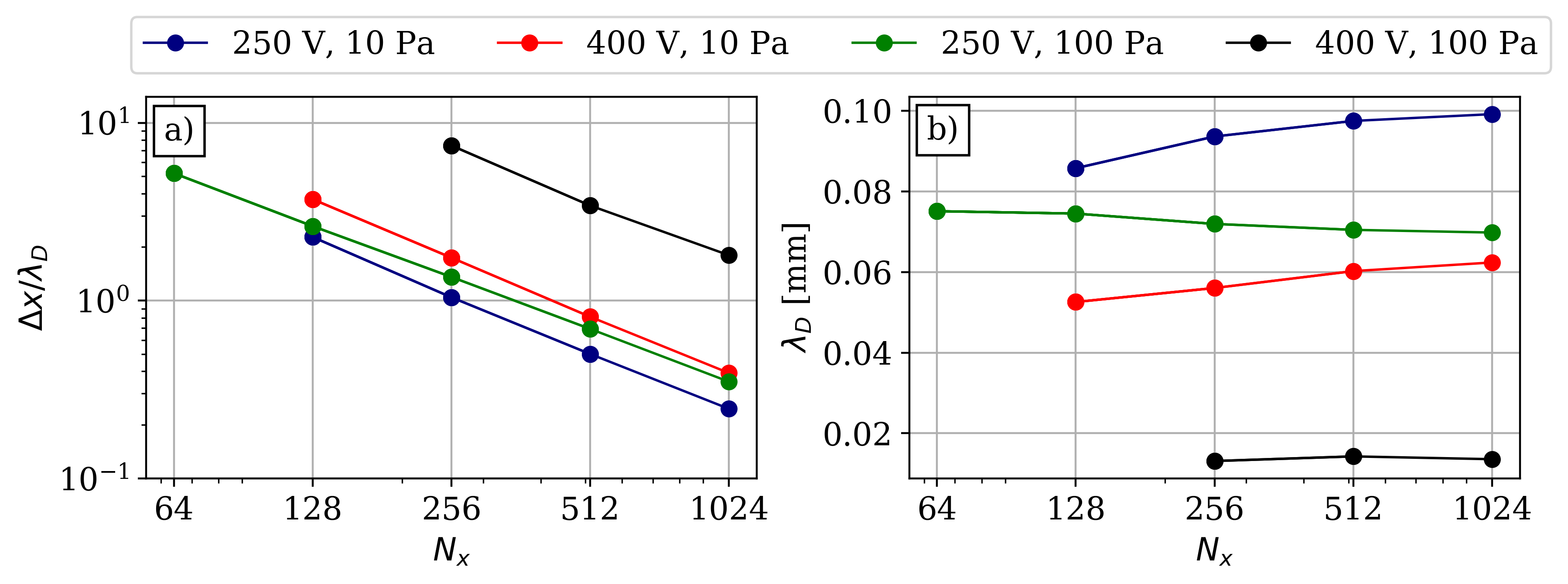}
    \caption{Ratio of the grid resolution, $\Delta x$, and the Debye-length, $\lambda_{\rm D}$, (a) and the value of the Debye-length (b) as a function of the number of grid points, $N_x$.}
    \label{fig:Lambda}
\end{figure*}

Consequently, we discuss only the condition requiring the resolution of the Debye-length. Figure~\ref{fig:Lambda}(a) shows the ratio between the grid resolution, $\Delta x$ and the Debye-length, $\lambda_{\rm D}$ together with the actual values of the Debye-length in panel (b) as a function of the number of grid points, $N_x$ for the different cases considered. Note, that as the plasma density peaks in the bulk region and the electron temperature (mean electron energy) is minimal in the center, the Debye-length, despite having a profound spatial distribution, is always minimal in the center of the discharge, as $\lambda_{\rm D}^2 \propto T_{\rm e}/n_{\rm e}$. For this reason, throughout this article, we refer to the time-averaged value at the center of the discharge as the Debye-length, characterizing each case investigated here. For the low pressure cases, where the transport is non-local, and electrons fly greater distances without collisions than at higher pressure, an accurate representation of the electric field is important. In this case, only the two highest $N_x$ cases were found to be converged in terms of density as shown in fig.~\ref{fig:10PX}(a,b). In panel (b) we see, that the Debye-length increases monotonically as a function of $N_x$, the reason being the decreased density as the resolution of the spatial grid gets finer. In contrast to this, the Debye-length for the high pressure cases seems to be constant, in accordance with the fact that the density does not change considerably except for the lowest value, $N_x=64$, as shown in fig.~\ref{fig:100PX}(a). Based on the data shown in panel (a) we can state that for cases where the transport is non-local, the Debye-length has to be resolved properly, as accurate results were only obtained in the case of $\Delta x\leq\lambda_{\rm D}$. This is intimately related to the resolution of the sheath and ambipolar electric fields: the spatial region of the sheath at low pressure is on the order of a few tens of the Debye-length~\cite{Lieberman}. If that length scale is not resolved properly, the sheath electric field is not resolved either, and, as already mentioned at figs.~\ref{fig:10PX} and \ref{fig:100PX}, the results can be inaccurate. The trend is similar for the high pressure case, albeit we can get accurate results in the case of $\phi_0=250$~V and $p=100$~Pa for $\Delta x/\lambda_{\rm D}=3$. Still, based on the $\phi_0=400$~V data, more accurate results could be achieved if the grid resolution is chosen such that $\lambda_{\rm D}\sim\Delta x$. To conclude, in this sense, the resolution of the Debye-length is an accuracy condition and not a stability condition, however, for the $\phi_0=400$~V, $p=100$~Pa case we found a numerically unstable regime for $\Delta x\sim O(10\lambda_{\rm D})$.

\begin{figure*}[htb]
    \centering
    \includegraphics[width=.8\textwidth]{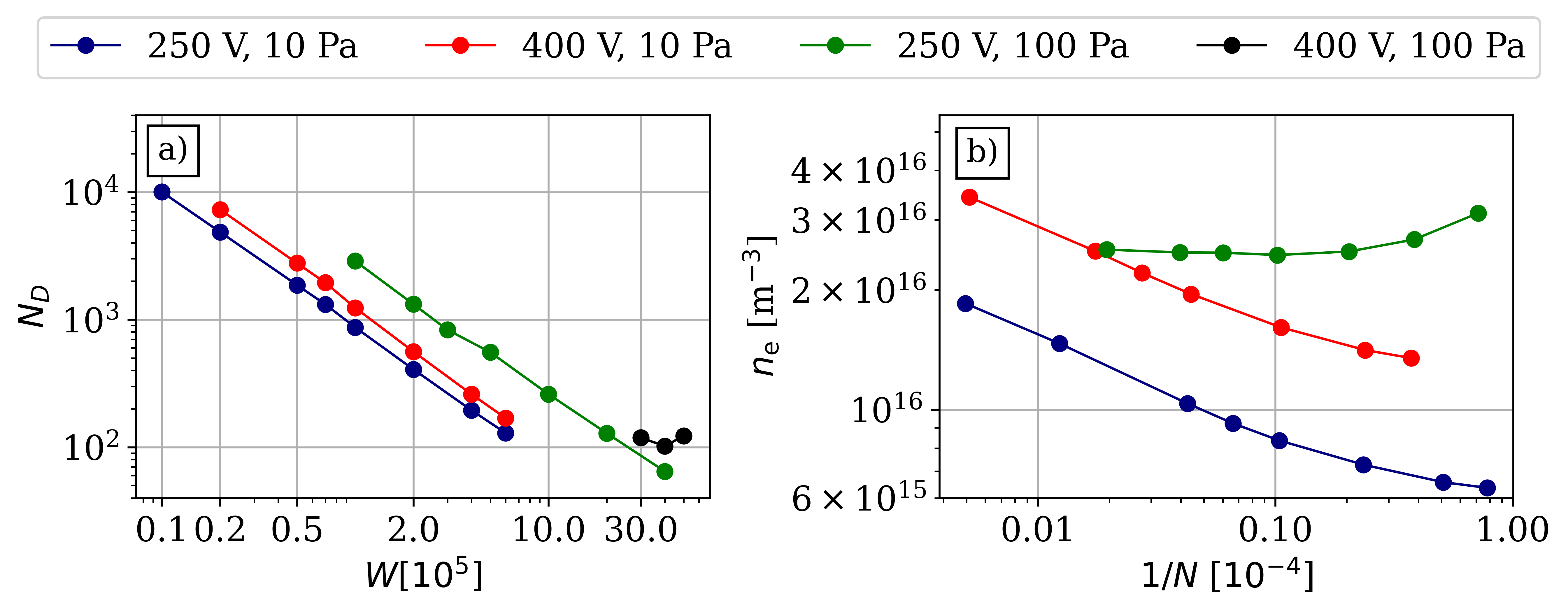}
    \caption{Number of particles within a Debye-sphere, $N_D$, as a function of the superparticle weight, $W$ (a) and mean electron density, $n_{\rm e}$, as a function of the inverse superparticle number, $1/N$ (b).}
    \label{fig:W}
\end{figure*}

Regarding the last stability criterion introduced in Sec.~\ref{sec:intro}, the requirement is to have ``enough'' particles in each numerical grid cell to provide a good statistical sampling of the charged particle distributions. In this formulation this clearly seems to be a numerical requirement, however, a physically meaningful version can be formulated by replacing the grid cell with the Debye-sphere. The Debye-sphere is the volume around a charge carrier in which its electrostatic effect persists, or in other words, any excess charge is screened by the plasma at a distance greater than the Debye-length. For this screening effect to work, however, the Debye-H\"uckel theory, developed originally for electrolytes \cite{DH}, assumes continuous charge density distributions. A large enough number of simulation particles is required to approximate this continuous distribution. As the resolution of the numerical grid is on the order of the Debye-length, usually, no distinction is necessary between the formulations of this criterion.

Figure~\ref{fig:W} shows the number of superparticles within a Debye-sphere, $N_D=N/\lambda_{\rm D}$ (where $N$ denotes the number of superparticles), as a function of the superparticle weight, $W$ (a) and the mean electron density, $n_{\rm e}$ as a function of the inverse superparticle number, $1/N$ (b). The corresponding criteria for the particle weight (or, equivalently, the number of particles) is, that there be ``many'' particles within one Debye-sphere (or, in our case, one Debye-length, $\lambda_{\rm D}$). As mentioned before, this is needed in order for the mean-field assumption of the PIC/MCC method to be applicable. The requirement of $N_D\gg1$, however, does not, by itself, determine the numerical value of $N_D$. For the $\phi_0=250$~V, $p=100$~Pa case, as shown in fig.~\ref{fig:W}(a,b), the density reaches a converged value in case of $N_D>200$. However, for the low pressure case, as already shown in fig.~\ref{fig:10PW}, increasing the number of particles, or, equivalently, decreasing the particle weight, increases the density, which, seemingly, does not converge to an asymptotic value, but increases indefinitely in the parameter range studied here. A discussion on possible reasons, consequences, and suggested extension of the model that could correct this weakness was discussed above. Based on the high pressure results we can conclude that this requirement is an accuracy rather than a stability criterion and that $N_D\approx200$ is large ``enough'' to provide accurate results.

\section{Conclusions}

The numerical stability conditions of particle in cell simulations have been investigated using a 1d3v electrostatic PIC/MCC code. The basic numerical parameters investigated were the number of timesteps, $N_t$, the number of grid points, $N_x$, the number of superparticles through the superparticle weight factor, $W$, as well as the electron reflection coefficient, $R$, and the ion induced secondary electron emission coefficient, $\gamma$, where the latter two describe the model of the electrode surfaces. Altogether four cases were chosen, with low/high pressure ($p=10$~Pa and 100~Pa) and low/high voltage amplitude ($\phi_0=250$~V and 400~V) using a single frequency waveform of $f=13.56$~MHz. Careful attention was paid to the distinction between stability and accuracy conditions, whereby the first determines the parameter range where converged results are possible at all, while the latter defines conditions where these converged results are numerically precise. The primary physical quantities investigated were the mean electron density together with the electron energy probability function (EEPF).

By changing the number of timesteps, and thus the minimal time resolution of the simulation ($\Delta t$), the density was found to increase as a function of $N_t$ for both high and low pressures. The reason for this was that when the resolution of the temporal grid is decreased, the value of the collision probability increases, which leads to ``collision misses'' and thus a decrease in the number of electron impact ionization and thus the density itself. Although, at low pressure, the differences in the density were found to be within $\approx10\%$, in the high pressure case the difference was found to be as large as a factor of 2. This also signifies the importance of the value of the collision probability, which, in this sense was found to be an accuracy condition. The condition, that the timestep has to resolve the inverse of the plasma frequency ($\omega_{\rm p,e}\Delta t<2$), however, was found to be a stability condition, as (i) numerical instabilities occurred for $\omega_{\rm p,e}\Delta t>2.2$ and (ii) stable results only changed by a factor of 2 compared to the 2 orders of magnitude difference between the lowest and highest values of $\omega_{\rm p,e}\Delta t$. 

Increasing the number of grid points, $N_x$ leads to the decrease of the density at low pressure. The reason for this is the inaccurate resolution of the deceleration field for the electrons, which thus can reach higher energies and contribute to ionization to a larger extent. For the high pressure case, the number of grid points was not as crucial, as, after a relatively low threshold value, the density was found to change relatively little with increasing $N_x$. The reason for this is the local transport at this high pressure, and thus the fact, that ionization only happens in the vicinity of the sheath edges. The condition, that the spatial grid ($\Delta x$) should resolve the Debye-length ($\lambda_{\rm D}$) was thus found to be both a stability and an accuracy condition: for values of $\Delta x\sim O(10\lambda_{\rm D})$, the simulations were found to be unstable, but a low $(\Delta x / \lambda_{\rm D})$ is required (at least at low pressure) to get accurate results. In summary, the number of grid points was found to be less important at higher pressures. Similar conclusions were drawn for the electron reflection coefficient, $R$: increasing this parameter increased the density at low pressure, while it had a very small effect on the density at high pressure. The reason for this can be found in the local nature of the transport at high pressure. However, the secondary electron emission coefficient, $\gamma$ was found to be important irrespective of the pressure.

Changing the number of superparticles, or, equivalently, the superparticle weight, had different effects at low/high pressure: at low pressure the density increased as the number of superparticles increased. The reason for this is the reductions of artificial isotropization (numerical heating), which is strong at low particle numbers, where the local fluctuations (i.e. deviations from average value, the basic assumption of the mean-field description of PIC) leads to a Maxwellian EEPF. This results in the depletion of the high energy tail of the EEPF, which leads to a decreased ionization and thus a decreased density. However, by increasing the number of superparticles (up to $2\cdot10^6$), no saturation was found for the density. As the increase of the number of superparticles resulted in an increased number of low energy electrons, our hypothesis is, that Coulomb-collisions (which are absent in our model) could stabilize this trend. A completely opposite behavior was observed for the high pressure cases. Here the density decreased as the number of superparticles increased. The reason for this is the local transport and the corresponding numerical heating of the local fluctuations in the electric field. However, after a certain threshold value (in our case $\approx 10^5$ particles) the value of the density converged. In the case of $p=100$~Pa, for $N_D>200$ saturation was achieved, while at low pressure the density kept increasing even for $N_D>10^4$. The condition for the number of particles is found to be an accuracy condition, as no unstable regime was found in the parameter range considered.

Although the change in the numerical parameters -- especially in the case of the superparticle weight -- lead to significant differences in the electron density, it was shown that this is not the case for the mean ion flux, and the flux-energy distribution of charged particles at the electrodes, which are far less susceptible to these changes that other physical parameters of the electrons.

\ack
This work was funded by the \'UNKP 20-3 New National Excellence Program of the Ministry for Innovation and Technology from the source of the National Research, Development and Innovation Fund, and by the German Research Foundation in the frame of the project, ``Electron heating in capacitive RF plasmas based on moments of the Boltzmann equation: from fundamental understanding to knowledge based process control'' (No. 428942393), and by the Hungarian Office for Research, Development, and Innovation via grant K-134462.

\section*{References}
\providecommand{\noopsort}[1]{}\providecommand{\singleletter}[1]{#1}%
\providecommand{\newblock}{}

\end{document}